\documentclass[12pt]{article}
\pdfoutput=1
\usepackage{jheppub}
\usepackage[utf8]{inputenc}
\usepackage{amsfonts,amsmath}

\usepackage{ifthen}
\usepackage{enumerate}
\usepackage{amsmath}
\usepackage{amssymb}
\usepackage[mathscr]{eucal}
\usepackage[errorshow]{tracefnt}
\usepackage{amsmath}
\usepackage{slashed}
\usepackage{amssymb}
\usepackage{array}
\usepackage{amsthm}
\usepackage{eucal}
\usepackage{epsf}
\usepackage{epsfig}
\usepackage{psfrag}
\usepackage{multirow}
\usepackage{wasysym} 
\usepackage{tikz-cd} 
\usepgfmodule{shapes}
\usetikzlibrary{backgrounds, arrows,calc,shapes,decorations.pathreplacing, decorations.markings, automata,positioning}

\usetikzlibrary{arrows,calc}
\usepackage{relsize}

\newcommand{\be}{\begin{equation}}
\newcommand{\ee}{\end{equation}}
\newcommand{\bea}{\begin{eqnarray}}
\newcommand{\eea}{\end{eqnarray}}
\newcommand{\ba}{\begin{array}}
\newcommand{\ea}{\end{array}}
\newcommand{\bpic}{\begin{tikzpicture}}
\newcommand{\epic}{\end{tikzpicture}}

\def\Tr{\rm Tr}
\newcommand{\tr}{\text{tr}\,}

\newcommand{\M}{{\mathfrak M}}

\newcommand{\Llra}{\Longleftrightarrow}

\newcommand\qt{\tilde q}

\newcommand\bt{\tilde b}
\newcommand\pt{\tilde p}

\newcommand\Qt{\tilde Q}

\newcommand{\CA}{{\mathcal A}}

\newcommand{\CN}{{\mathcal N}}

\newcommand{\CT}{{\mathcal T}}

\newcommand{\CW}{{\mathcal W}}

\newcommand{\cA}{\mathcal{A}}

\newcommand{\cN}{\mathcal{N}}

\newcommand{\cW}{\mathcal{W}}

\def\b{\beta} 
\def\g{\gamma}

\newcommand\s{\sigma}

\def\Bt{\tilde{B}}

\title{Sequential deconfinement in $3d$ $\cN\!=\!2$ gauge theories}

\author[1]{Sergio Benvenuti}
\author[2]{Ivan Garozzo}
\author[3,4]{Gabriele Lo Monaco}

\affiliation[1]{INFN Sezione di Trieste, via Bonomea 265, 34136 Trieste, Italy}
\affiliation[2]{Dipartimento di Fisica, Universit\`a di Milano-Bicocca $\&$ INFN, Sezione di Milano-Bicocca,
I-20126 Milano, Italy}
\affiliation[3]{Institut de Physique Th\'eorique, Universit\'e Paris Saclay, CEA, CNRS, \\ Orme des Merisiers, 91191 Gif-sur-Yvette CEDEX, France}
\affiliation[4]{Department of Physics, Stockholm University, AlbaNova, 10691 Stockholm, Sweden}

\abstract{We consider $3d$ $\cN\!=\!2$ gauge theories with fundamental matter plus  a single field in a rank-$2$ representation. Using iteratively a process of ``deconfinement'' of the rank-$2$ field, we produce a sequence of Seiberg-dual quiver  theories. We detail this process in two examples with zero superpotential: $Usp(2N)$ gauge theory with an antisymmetric field and $U(N)$ gauge theory with an adjoint field. The fully deconfined dual quiver has $N$ nodes, and can be interpreted as an Aharony dual of theories with rank-$2$ matter. All chiral ring generators of the original theory are mapped into gauge singlet fields of the fully deconfined quiver dual.}

\emailAdd{benve79@gmail.com}
\emailAdd{i.garozzo@gmail.com}
\emailAdd{gabriele.lomonaco@ipht.fr}

\begin{document}

\maketitle

\section{Introduction and results}
The fascinating phenomenon of infrared dualities seems ubiquitous in strongly coupled gauge theories living in $d\leq 4$ dimensions. In the special subset of supersymmetric theories with $4$ supercharges, many examples of such dualities have been discovered, starting from \cite{Seiberg:1994pq}. Impressive checks of the dualities are possible: matching of the infrared global symmetry, of the chiral ring and of various supersymmetric partition functions.

In the case of $3d$ $\cN\!=\!2$ gauge theories \cite{deBoer:1997ka, deBoer:1997kr, Aharony:1997bx, Aharony:1997gp}, the simplest and paradigmatic examples are the Aharony dualities \cite{Aharony:1997gp}, which relate a pair of theories with a single gauge group. $Usp(2N)$ with $2f$ flavors is dual to $Usp(2f-2N-2)$ with $2f$ flavors, while $U(N)$ with $(F,F)$ flavors is dual to $U(F-N)$ with $(F,F)$ flavors. All chiral ring generators, both mesons and monopoles, of the electric theory are mapped into gauge singlet fields in the magnetic theory.

In this paper we consider $3d$ $\cN\!=\!2$ gauge theories with matter content consisting of an arbitrary number of fundamental flavors and a single field in a rank-$2$ representation. A rank-$2$ field can sometimes be \emph{deconfined}, as shown in the early days of Seiberg dualities in $4d$ $\cN=1$ models \cite{Berkooz:1995km,Pouliot:1995me,Terning:1997jj}. In $3d$ $\cN=2$ the story is similar, with the difference that in $3d$ monopole operators play a crucial role. Examples studied in $3d$ $\cN=2$ include \cite{Nii:2016jzi, Pasquetti:2019uop, Pasquetti:2019tix}. In particular, the approach of Pasquetti-Sacchi 
is particularly interesting, since it allows to find the dual of $U(N)$ with one adjoint field and one flavour \cite{Pasquetti:2019uop} and the one for $U(N)$ with one adjoint field and $k+1$ flavours 
\cite{Pasquetti:2019tix} starting from free field correlators in $2d$ Liouville CFT. These results have also been uplifted to four dimensions and are related to the compactification of rank-$Q$ E-string theory on a torus with flux \cite{Pasquetti:2019hxf}, and subsequently lead to the discovery of an analogue of $3d$ mirror symmetry for $4d$ $\mathcal N=1$ theories \cite{Hwang:2020wpd}. 

We use a process of ``sequential deconfinement'' in order to find a quiver dual of a $3d$ $\CN=2$ gauge theory with gauge group $G$ and a single rank-$2$ matter field. See \cite{Benini:2011mf, Amariti:2014lla, Benvenuti:2016wet, Amariti:2017gsm, Benvenuti:2017kud,Benvenuti:2017bpg, Zenkevich:2017ylb, Aprile:2018oau, Amariti:2019pky, Pasquetti:2019uop, Pasquetti:2019tix} for recent works in $3d$ $\cN=2$ quivers. Let us also mention that a similar technique has been recently exploited in the context of $2d$ $(0,2)$ supersymmetric field theories to find duals of $Usp(2N)$ gauge theory with one antisymmetric chiral, four fundamental chirals and $N$ Fermi singlets \cite{Sacchi:2020pet}. 
The iterative (or sequential) application of Seiberg dualities in quiver gauge theories played a crucial role in various recent works concerning $3d$ $\CN=2$ QFT's, see for instance \cite{Benvenuti:2017kud, Benvenuti:2017bpg, Aprile:2018oau, Pasquetti:2019uop, Pasquetti:2019tix}. 

In this paper we deconfine using Aharony duality \cite{Aharony:1997gp} or its variants with monopole superpotential \cite{Benini:2017dud} or Chern-Simons interactions \cite{Willett:2011gp, Benini:2011mf}.

The main complication in the process is given by the supersymmetric monopole operators \cite{Borokhov:2002ib, Borokhov:2002cg}. Monopole operators appear in the superpotential, both linearly and through flipping-type interactions. Moreover it is important at each step to keep track of the mapping of the monopoles across the dualities. Hence we need to control the monopoles in $3d$ $\CN=2$ quivers, for we which we use the results of \cite{Benvenuti:2020wpc, Pasquetti:2019uop, Pasquetti:2019tix}.

\subsection*{Results}

In this paper we focus on two examples: $Usp(2N)$ with an antisymmetric and  $U(N)$ with an adjoint. Let us state the final results.

For unitary-symplectic gauge group, we find in Section \ref{asymmSp} that $Usp(2N)= Sp(N)$ gauge theory with antisymmetric, $2f$ flavors and zero superpotential, is dual to a quiver theory with $N$ nodes:
  \be \label{RESSp} \bpic  \path (-6,0) node[blue](x0) {\small$Sp(\!N(f\!-\!2)\!)$} --    (-2,0) node[blue](x1) {\small$Sp(\!(\!N\!-\!1\!)\!(f\!-\!2)\!)$} -- (1,0) node[blue](x2) {$\ldots$} -- (3,0) node[blue](x3) {\small{$Sp(\!f\!-\!2\!)$}}  -- (-6,2.5) node[rectangle,draw](x6) {$\,2f\,$};
  \path (1,3) node {$ \CW= \sum_{i=1}^{N-1} \g_i \M^{0, \bullet^{N-i}, 0^{i-1}} +\sum_{i=1}^{N} \s_i \M^{\bullet^{N-i+1}, 0^{i-1}}  + $}-- (-3,2.3) node[right] {$  +  \sum_{i=1}^{N} M_i \tr(p b_{N-1} \ldots b_{i} b_{i}  \ldots b_{N-1} p) +$}-- (-3,1.6) node[right]  {$ + \sum_{i=1}^{N-1} ( \tr(A_i b_i b_i) + \tr(A_i b_{i-1} b_{i-1}) + a_i \tr(b_i b_i)) $};
\draw [-] (x1) to (x0); 
\draw [-] (x1) to (x2); 
\draw [-] (x0) to (x6); 
\draw [-] (x2) to (x3); 
\draw [-] (x6) to[out=45, in=0] (-6,3.1) to[out=180,in=135] (x6);
\draw [-] (x6) to[out=45, in=0] (-6,3.3) to[out=180,in=135] (x6);
\draw [-] (x6) to[out=45, in=0] (-6,3.7) to[out=180,in=135] (x6);
\node at (-5,3.2){$M_i$};
\node at (-6,3.5){$\ldots$};
\draw [-] (x1) to[out=-45, in=0] (-2,-0.8) to[out=180,in=-135] (x1);
\node at (-1.1,-0.6){$A_{N-1}$};
\draw [-] (x3) to[out=-45, in=0] (3,-0.8) to[out=180,in=-135] (x3);
\node at (3.6,-0.6){$A_1$};
\node[above right] at (-0.5,-0.1){$b_{N-2}$};
\node[above right] at (-4.8,-0.1){$b_{N-1}$};
\node[above right] at (1.5,-0.1){$b_1$};
\node at (-5.8,1.4){$p$}; 
  \epic\ee 
  
Similarly, for unitary gauge group, we find in Section \ref{adjU} that $U(N)$ with an adjoint, $F$ fundamentals, $F$ antifundamentals, $\cW=0$, is dual to the following quiver
\be\label{RESU}
\begin{tikzpicture}[baseline]
\scalebox{1.1}{
\tikzstyle{every node}=[font=\footnotesize]
\node[draw=none] (g1) at (0,0) {$U(N(F-1))$};
\node[draw=none] (g2) at (4,0) {$U((N-1)(F-1))$};
\node[draw=none] (g3) at (7,0) {$\cdots$};
\node[draw=none] (g4) at (9.3,0) {$U(F-1)$};
\node[draw, rectangle] (f1) at (0,2.5) {$F$};
\draw[black,solid] (g2) edge [out=-45,in=-135,loop,looseness=4] (g2);
\node[draw=none] at (4.1, -1.2) {$\Phi_{N-1}$} ;
\draw[black,solid] (g4) edge [out=-45,in=-135,loop,looseness=4] (g4);
\node[draw=none] at (9.4, -1.2) {$\Phi_{1}$} ;
\draw[transform canvas={yshift=2.5pt},->] (g1) to (g2) ;
\draw[transform canvas={yshift=-2.5pt},<-] (g1) to (g2) ;
\node[draw=none] at (1.8, -0.5) {$b_{N-1}, \bt_{N-1}$} ;
\draw[transform canvas={yshift=2.5pt},->] (g2) to (g3) ;
\draw[transform canvas={yshift=-2.5pt},<-] (g2) to (g3) ;
\node[draw=none] at (6.2, -0.5) {$b_{N-2}, \bt_{N-2}$} ;
\draw[transform canvas={yshift=2.5pt},->] (g3) to (g4) ;
\draw[transform canvas={yshift=-2.5pt},<-] (g3) to (g4) ;
\node[draw=none] at (8, -0.5) {$b_{1}, \bt_{1}$} ;
\draw[transform canvas={xshift=2.5pt},->] (g1) to (f1) ;
\draw[transform canvas={xshift=-2.5pt},<-] (g1) to (f1) ;
\draw[black,solid] (f1) edge [out=45,in=135,loop,looseness=2] (f1);
\draw[black,solid] (f1) edge [out=45,in=135,loop,looseness=3] (f1);
\draw[black,solid] (f1) edge [out=45,in=135,loop,looseness=7] (f1);
\node[draw=none] at (0.03, 3.25) {$\cdots$} ;
\node[draw=none] at (0.9, 3.25) {$M_i$} ;
\node[draw=none] at (4.3, 3)  {$ \CW= \sum_{i=1}^{N-1}  \M^{0^{N-i}, - , 0^{i-1}} + $};
\node[draw=none] at (6.6, 2.4)  {$ + \sum_{i=1}^{N-1} \g_i \M^{0, +^{N-i}, 0^{i-1}} +\sum_{i=1}^{N} \s^{\pm}_i \M^{\pm^{N-i+1}, 0^{i-1}}  + $};
\node[draw=none] at (5.72, 1.8)  {$  +  \sum_{i=1}^{N} M_i \tr(\pt \bt_{N-1} \ldots \bt_{i} b_{i}  \ldots b_{N-1} p) +$};
\node[draw=none] at (6.49, 1.2) {$ + \sum_{i=1}^{N-1} ( \tr(\Phi_i \bt_i b_i) + \tr(\Phi_i \bt_{i-1} b_{i-1}) + \phi_i \tr(\bt_i b_i)) $};
}
\end{tikzpicture}
\ee

In the main text we explain the notation and derive these dualities, together with the mapping of the chiral ring generators.

The dualities  \eqref{RESSp} and \eqref{RESU} are valid for vanishing superpotential in the electric single-node theory, so it is possible to turn on any superpotential and obtain new duals. Similarly, turning on real masses it is possible to obtain duals of theories with non zero Chern-Simons level. We explore various such deformed dualities in the main text.

One noteworthy feature of the dualities  \eqref{RESSp} and \eqref{RESU} is that all chiral ring generators, both mesons and monopoles, of the electric theory are mapped into gauge singlet fields in the magnetic theory. So in this sense they are a natural generalization of Aharony dualities to the case with a single rank-$2$ matter field.

\subsection*{Further directions}

A similar sequential deconfinement procedure can be worked out for theories involving orthogonal gauge groups and/or rank-$2$ matter in a symmetric representation. We study such a process in \cite{sawadjoint}. The deconfined quivers alternate a symplectic and an orthogonal group. Moreover, it turns out that the quivers display a \emph{saw} structure.

$3d$ $\cN=2$ gauge theories with a single gauge group, rank-$2$ matter $\Phi$, fundamentals and superpotential $\cW=\Tr(\Phi^{k+1})$ are known to admit a single node dual of Kutasov-Schwimmer type, that is the dual has a single node, a tower of gauge singlets and a superpotential term $\cW=\Tr(\tilde{\Phi}^{k+1})$ \cite{Kim:2013cma, Nii:2014jsa, Hwang:2018uyj, Amariti:2018wht, Nii:2019qdx, Amariti:2020xqm}. Such dualities appear different from the dualities discussed in this paper, which have $\cW=0$ on the l.h.s. and a linear quiver on the r.h.s. It would be interesting to investigate a possible relation between the Kutasov-Schwimmer type dualities and our sequential deconfinement procedure.

Another possible direction, which was one of the main motivation for this study, is to extend these results to $3d$ theories with $\cN=1$ supersymmetry and rank-$2$ matter (see \cite{Bashmakov:2018wts, Benini:2018umh, Eckhard:2018raj, Gaiotto:2018yjh, Benini:2018bhk, Choi:2018ohn, Fazzi:2018rkr, Rocek:2019eve, Aharony:2019mbc, Bashmakov:2018ghn, Bashmakov:2019myq,   Sharon:2020xod} for recent results in $3d$ $\cN=1$ gauge theories).
Very little is known on the dynamics of rank-2 matter for $\mathcal N=1$ theories. We hope that a story similar to the one in the present paper is valid in the $3d$ $\cN=1$ realm, which might be at midway between the $\cN=2$ and the non-supersymmetric case \cite{Gomis:2017ixy, Choi:2018tuh, Choi:2019eyl}. In particular, the IR dynamics of non-supersymmetric theories with two real adjoint fields, unveiled in \cite{Choi:2019eyl}, displays an intricate \emph{duality chain} reminiscent of the $\cN=2$ sequential deconfinement.

\section{A sequence of duals for $Usp(2N)$ with an antisymmetric}\label{asymmSp}
In this section we find dual descriptions of $Usp(2N)=Sp(N)$ ($Sp(1)=SU(2)$) with a field $\cA$ in the traceless antisymmetric representation of $Sp(N)$ and $2f$ complex flavors $Q_i$, $\cW=0$. $Usp(2N)$ theories have been recently studied in \cite{Amariti:2018wht, Benvenuti:2018bav, Nii:2019ebv}.
We consider $f \geq 3$. If $f=1$, the theory does not have a supersymmetric vacuum. If $f = 2$, the fully deconfined dual is a Wess-Zumino model, see  \cite{Amariti:2018wht, Benvenuti:2018bav}.

We find a total of $2N$ dual theories, that are quivers with a number of nodes ranging from $1$ to $N$, the most natural one being the fully deconfined dual, with $N$ nodes. 

In each model we describe the chiral ring, giving the list of the chiral ring generators and their global symmetry quantum numbers. As usual in $3d$ gauge theories, we need to pay special attention to the monopole operators. 

We first consider the case of vanishing Chern-Simons interactions, with this result, it will be easy to turn on a real mass deformation and hence a Chern-Simons term in section \ref{addCS}.

We start with theory $\CT_1$, that is $Sp(N)$ with a traceless antisymmetric field $\cA$ and $2f$ complex flavors $Q_i$. We take the  superpotential to be vanishing. Using the standard quiver notation for theories with four supercharges, $\CT_1$ reads
\be \label{T1} \bpic  
\path (-5,0) node{$\CT_1:$} -- (-3,0) node[blue](x1) {$Sp(\!N\!)$} -- (-0.5,0) node[rectangle,draw](x2) {\small{$2f$}};
\draw [-] (x1) to (x2); 
\draw [-] (x1) to[out=-45, in=0] (-3,-0.7) to[out=180,in=225] (x1);
\node at (3,0){$ \CW= 0$};
\node[below right] at (-3,-0.5){$\cA$};
\node at (-1.6,0.3){$Q_i$}; \epic  \ee

The chiral ring is generated by the (dressed) mesons $\tr(Q_i \cA^l Q_j)$, $l=0,\ldots,N-1$, the powers of the antisymmetric traceless field $\tr(\cA^j)$, $j=2,\ldots,N$, and  the (dressed) monopoles $\{\M_{\cA^k}\}$, $k=0,1,\ldots,N-1$. In terms of the $R$-charges of the elementary fields $Q_i$ and $\cA$, which we denote $r_F$ and $r_\cA$, the $R$-charge of the basic, undressed, monopole $\M$ is
\be R[\M]_{\CT_{1}} = 2f(1-r_F) + (2N-2)(1-r_\cA)-2N= 2f(1-r_F) -(2N-2)r_\cA-2 \ee

\subsection{Deconfine and dualize: first step}\label{step1Sp}
We now use the \emph{confining} duality\footnote{This is a variation of a duality introduced by Aharony in \cite{Aharony:1997gp}:
\be
\ba{c} Sp(N-1) \, \text{w/ $2N$ chiral flavors} \\ \cW= 0 \ea
     \qquad      \Llra \qquad
\ba{c}   \text{Wess-Zumino w/} \, 2N \times 2N  \, \text{antisymmetric  }\\ \text{matrix of chiral fields $A$, and a singlet $\s$ }\\ \cW= \s\, \text{Pfaff}(A)  \ea \ee
In this duality the monopole is mapped to $\s$ ($\M \leftrightarrow \s$), so if we flip the monopole in the l.h.s. with a gauge singlet $\g$, on the r.h.s. we obtain a superpotential term $\s \g$, so $\s$ and $\g$ become massive, integrating them out the superpotential becomes zero and we obtain the duality  \eqref{decSp}.}
\be  \label{decSp}
\ba{c} Sp(N-1) \, \text{w/ $2N$ chiral flavors}\, q_i \\ \cW= \g \, \M \ea
   \qquad        \Llra   \qquad
\ba{c}   N(2N-1) \,\, \text{free chirals}\,A_{ij}\\ \text{antisymmetric of}\,\,SU(2N)  \ea \ee
In this duality the chiral ring generators map as $\tr(q_i q_j) \leftrightarrow A_{ij}$ (the monopole $\M$ and the singlet $\g$ are zero in the chiral ring).

Starting from theory $\CT_{1}$, we \emph{deconfine the antisymmetric field} into a two-node quiver theory. That is we consider theory $\CT_{1'}$:
  \be \bpic  \path (-3.5,1.5) node {$\CT_{1'}:$}   (-3,0) node[blue](x1) {$Sp(\!N\!-\!1\!)$} -- (0,0) node[blue](x2) {$Sp(\!N\!)$}  -- (0,2) node[rectangle,draw](x6) {$\,2f\,$}  -- (4,1) node {$\CW= \gamma \, \M^{\bullet,0}+ \b \tr(\bt \bt)$};
\draw [-] (x1) to (x2); 
\draw [-] (x2) to (x6);
\node[above right] at (-1.7,0){$\bt$};
\node at (-0.2,1){$Q$}; 
  \epic\ee 
Applying the duality  \eqref{decSp} to the left node of $\CT_{1'}$, the node $Sp(N-1)$ \emph{confines} and one readily obtains $\CT_{1}$. So $\CT_1$ and $\CT_{1'}$ are dual. We introduced the gauge singlet field $\b$ field so that $\cA$ in  $\CT_{1}$ is traceless. The mapping of the $R$-charges between $\CT_{1}$ and $\CT_{1'}$ is simply $r_Q=r_F, r_\cA=2 r_{\bt}$.

In linear quivers made of $Sp$ gauge groups, we denote by $\M^{0,\bullet,0,0,\ldots}$ monopoles with non-zero minimal flux in the nodes with $\bullet$ and zero flux in nodes with $o$.

In theory $\CT_{1'}$, $\M^{\bullet,0}, \gamma, \beta$ are zero in the chiral ring: $\M^{\bullet,0}$ is set to zero by the F-terms of $\g$. $\g$ and $\b$ cannot take a vev because of quantum generated superpotentials, so we expect them to be zero in the chiral ring\footnote{This argument leaves the logical possibility that they are nilpotent operators, but we do not expect this possibility to be realized.}. The monopoles $\M^{0,\bullet}$ and $\M^{\bullet,\bullet}$ are instead non-zero the chiral ring, their $R$-charges are
\be R[\M^{0,\bullet}]_{\CT_{1'}} = 2f(1-r_Q) + (2N-2)(1-r_{\bt})-2N \ee
\be R[\M^{\bullet,\bullet}]_{\CT_{1'}} = 2f(1-r_Q) + (2N-2+2N-2)(1-r_{\bt}) - 2(N-1)-2N = 2f(1-r_Q)-4(N-1)r_{\bt}-2\ee
and (using that $r_Q=r_F, r_\cA=2 r_{\bt}$) are equal to $ R[ \{\M_{\cA^{N-1}}\}]_{\CT_{1}}
$ and $R[\M]_{\CT_{1}}$, respectively.

The basic monopole $\M$ in $\CT_1$ maps to the 'extended' monopole $\M^{\bullet,\bullet}$ in $\CT_{1'}$. We will give the full map of the chiral ring generators in \eqref{map1}. As explained in \cite{Benvenuti:2020wpc}, the monopole $\M^{\bullet,\bullet}$  in $\CT_{1'}$ can be dressed with the square of bifundamental field, that is $\bt \bt$, in same way that $\M$ in $\CT_1$ can be dressed with the antisymmetric field $\cA$.

The next step is to dualize the right node $Sp(N)$ in $\CT_{1'}$. We  use the Aharony duality \cite{Aharony:1997gp}
\be \ba{c} Sp(N) \,\, \textrm{w/} \,\, 2F \,\, \textrm{flavors},\\ \CW=0 \ea
 \Longleftrightarrow 
 \ba{c}Sp(F-N-1)\,\, \textrm{w/} \,\, 2F \,\, \textrm{flavors}\,\, p_i,\\ \CW= A^{ij} \tr(p_i p_j) + \s \M \ea \ee
in the quiver $\CT_{1'}$ and obtain $\CT_{2}$:
  \be\label{CT2Sp} \bpic  \path (-4,2.5) node {$\CT_{2}:$}   (-3,0) node[blue](x1) {$Sp(\!N\!-\!1\!)$} -- (0,0) node[blue](x2) {$Sp(\!f\!-\!2\!)$}  -- (-1.5,2.5) node[rectangle,draw](x6) {$\,2f\,$}  -- (3,2) node {$ \CW=  \gamma \M^{\bullet,\bullet} + \s \M^{0,\bullet} +$} -- (3.8,1.4) node {$ + \tr(b \phi b)+ \tr(b q p)  + M \tr(q q)$};
\draw [-] (x1) to (x2); 
\draw [-] (x2) to (x6); 
\draw [-] (x1) to (x6); 
\draw [-] (x1) to[out=-45, in=0] (-3,-1) to[out=180,in=225] (x1);
\draw [-] (x6) to[out=45, in=0] (-1.5,3.5) to[out=180,in=135] (x6);
\node[below right] at (-3,-0.6){$\phi$};
\node at (-0.8,3){$M$};
\node[above right] at (-1.8,0.3){$b$};
\node at (-1.8,1.5){$p$}; 
\node at (-1.2,1.4){$q$}; 
  \epic\ee 
We decomposed the dual mesons into the two fields $\phi$ and $p$. Because of the F-terms of the singlet $\b$, that we integrated out, the antisymmetric field $\phi$ is traceless. Notice that the monopole $\M^{\bullet,0}$ in $\CT_{1'}$ maps to $\M^{\bullet,\bullet}$ in $\CT_{2}$, \eqref{CT2Sp}; here we are applying the rules of  \cite{Benvenuti:2020wpc} for the mapping of monopole operators under dualities in quivers made of $Sp$ nodes.

The mapping between the $R$-charges  of theories $\CT_{1}$ and $\CT_2$ is 
\be\label{map1}  r_q = 1-r_F\,, \qquad r_p=r_\cA/2+r_F\,, \qquad r_{\phi}=r_\cA\,, \qquad r_b=1-r_\cA/2\,.\ee

The $R$-charges of the monopoles and of the flipping fields for the monopoles are
\be R[\M^{\bullet,0}]_{\CT_{2}} = (2f-4)(1-r_b) + 2f(1-r_p) + (2N-4)(1-r_\phi)-2N+2  \ee
\be R[\s]_{\CT_{2}} = 2- R[\M^{0,\bullet}]_{\CT_{2}}=2-((2N-2)(1-r_b)+2f(1-r_q)-(2f-4)) \ee
\be R[\gamma]_{\CT_{2}} = 2-((2N+2f-8)(1-r_b)+(2N-4)(1-r_\phi)+2f(2-r_q-r_p)-(2N+2f-6)) \ee
which, using \eqref{map1}, in terms of the $R$-charges of $\CT_1$, become
\be R[\M^{\bullet,0}]_{\CT_{2}} =2f(r_q)-(2N-2)r_\phi-2= R[\M]_{\CT_{1}}\ee
\be R[\s]_{\CT_{2}} = 2f(r_q)-(N-1)r_\phi-2  =R[\M]_{\CT_{1}} + (N-1)r_\phi =R[{\M_{\cA^{N-1}}}]_{\CT_{1}}\ee
\be R[\gamma]_{\CT_{2}} =  N r_\cA \ee

The mapping of the chiral ring generators of the three theories constructed so far, $\CT_1$, $\CT_{1'}$ and $\CT_2$ is
\be \label{map1}
\ba{c}\CT_1 \\
\tr( Q_i Q_j)  \\
\tr( Q_i \cA^J Q_j)  \\
\tr(\cA^J)\\
\tr(\cA^{N})\\
 \{\M_{\cA^J}\}\\
  \{\M_{\cA^{N-1}}\}
\ea
\qquad \Longleftrightarrow \qquad
\ba{c} \CT_{1'}\\
\tr( Q_i Q_j)  \\
\tr( Q_i (\bt \bt)^J Q_j)  \\
\tr((\bt \bt)^J)\\
\tr((\bt \bt)^N)\\
 \{\M^{ \bullet,\bullet}_{(\bt \bt)^J}\}\\
  \M^{0,\bullet}
\ea
\qquad \Longleftrightarrow \qquad
\ba{c}\CT_2\\
M_{ij} \\
 \tr(p_i \phi^{J-1} p_j)\\
\tr(\phi^J)\\
 \gamma \\
 \{\M^{ \bullet,0}_{\phi^J}\} \\
 \s
\ea
\qquad
\ba{l}
\,\\
\,\\
J=1,\ldots,N-1\\
J=2,\ldots,N-1\\
\\
J=0,1,\ldots,N-2\\
\,\ea
 \ee

It is possible to check the mapping of the dressed mesons using
\be 
R[\tr( Q_i (\bt \bt)^J Q_j)]_{\CT_{1'}} = 2 r_Q + 2J r_{\bt} = 2-2r_q +  r_{\phi}+(J-1) r_{\phi} =\ee
\be 2(2-r_q -r_b) +(J-1) r_{\phi}=  2 r_p + (J-1)r_{\phi} = R[\tr(p_i \phi^{J-1} p_j)]_{\CT_2} \ee

\subsection{Deconfine and dualize: second step}
We now repeat the same procedure. First we deconfine the antisymmetric traceless in $\CT_2$ \eqref{CT2Sp} into a bifundamental $\tilde{b}$ connected to a $Sp(N-2)$ node, introducing a flipping field $\g_2$ for the $Sp(N-2)$-monopole. The superpotential term  $\tr(b  \phi b)$ becomes $\tr(b \tilde{b} \tilde{b} b)$ and we get $\CT_{2'}$: 
  \be \bpic  \path  (-5.5,2) node {$\CT_{2'}:$} -- (-6,0) node[blue](x0) {\small$Sp(\!N\!-\!2\!)$} --    (-3,0) node[blue](x1) {\small$Sp(\!N\!-\!1\!)$} -- (0,0) node[blue](x2) {\small{$Sp(\!f\!-\!2\!)$}}  -- (-1.5,2.5) node[rectangle,draw](x6) {$\,2f\,$}  -- (3.6,2) node {$ \CW=  \g_1 \M^{\bullet,\bullet,\bullet} +\g_2 \M^{\bullet,0,0} + \s_1 \M^{0,0,\bullet}+ $}-- (4.8,1.3) node {$ +  g_2 \tr(\tilde{b}  \tilde{b})+\tr(b \tilde{b} \tilde{b} b) +\tr(b q p)  +\tr(q M_1 q) $};
\draw [-] (x1) to (x0); 
\draw [-] (x1) to (x2); 
\draw [-] (x2) to (x6); 
\draw [-] (x1) to (x6); 
\draw [-] (x6) to[out=45, in=0] (-1.5,3.5) to[out=180,in=135] (x6);
\node at (-0.7,3){$M_1$};
\node[above right] at (-1.8,0){$b$};
\node[above right] at (-4.8,0){$\tilde{b}$};
\node at (-1.9,1.4){$p$}; 
\node at (-1.2,1.4){$q$}; 
  \epic\ee 
Notice that the monopole $\M^{\bullet,\bullet}$ in $\CT_2$ is extended to $\M^{\bullet,\bullet,\bullet}$ in $\CT_{2'}$, while $\M^{0,\bullet}$ in $\CT_2$ is becomes $\M^{0,0,\bullet}$ in $\CT_{2'}$. This is agreement with the rules of \cite{Benvenuti:2020wpc}, since dualizing the leftmost node in $\CT_{2'}$ (and forgetting that the rank of the leftmost group becomes zero), the rule says that $\M^{\bullet,\bullet,\bullet}\rightarrow \M^{0,\bullet,\bullet}$ and $\M^{0,0,\bullet}\rightarrow \M^{0,0,\bullet}$.

Dualizing the central $Sp(N-1)$ node, which has a total of $2(N-2+f-2+f)$ flavors, and thus becomes a $Sp(N+2f-4-(N-1)-1=2f-4)$ with $3$ sets of flavors $p', b_1', b_2'$, we produce $6$ sets of Seiberg mesons $\hat{A}_{1}, \hat{\phi}, M_2, N_{13}, N_{1F}, N_{3F}$. The quiver becomes
  \be \bpic  
  \path (-6,0) node[blue](x0) {\small$Sp(\!N\!-\!2\!)$} --    (-3,0) node[blue](x1) {\small$Sp(\!2f\!-\!4\!)$} -- (0,0) node[blue](x2) {\small{$Sp(\!f\!-\!2\!)$}}  -- (-3,2.5) node[rectangle,draw](x6) {$\,2f\,$};
  \path (3.6,2.2) node {$ \CW=  \g_1 \M^{\bullet,\bullet,\bullet} +\g_2 \M^{\bullet,\bullet,0} + \s_1 \M^{0,\bullet,\bullet}+ \s_2 \M^{0,\bullet,0}+$}-- (4.8,1.6) node {$ + g_2 \tr(\hat{\phi}) + \tr(N_{13} N_{13}) +\tr(N_{1F} q)  +\tr(q q M_1 ) + $}-- (4.8,1) node {$  \hat{\phi} b'_2 b'_2 + \hat{A}_1 b'_1 b'_1 + M_2 p' p' + N_{13}b'_1 b'_2 + N_{1F}p' b_1' + N_{3F} p' b'_2 $};
\draw [-] (x1) to (x0); \node at (-1.7,-0.3){$b'_1$};
\draw [-] (x1) to (x2); \node[above right] at (-4.8,-0.1){$b'_2$};
\draw [-] (x2) to[bend right]  (x6); \node at (-1,1.4){$q$}; 
\draw [-] (x2) to[bend left]  (x6); \node at (-1.9,1.2){$N_{1F}$}; 
\draw [-] (x1) to (x6); \node at (-3.2,1.2){$p'$}; 
\draw [-] (x0) to (x6);  \node at (-4.2,1.2){$N_{3F}$};
\draw [-] (x0) to[out=-25, in=180] (-3,-0.9) to[out=0,in=205] (x2);\node at (-3,-1.2){$N_{13}$};
\draw [-] (x6) to[out=45, in=0] (-3,3.5) to[out=180,in=135] (x6);
\draw [-] (x6) to[out=45, in=0] (-3,3.3) to[out=180,in=135] (x6);
\node at (-1.8,3){$M_1,M_2$};
\draw [-] (x0) to[out=-45, in=0] (-6,-0.8) to[out=180,in=-135] (x0);
\node at (-5.4,-0.6){$\phi$};
\draw [-] (x2) to[out=-45, in=0] (0,-0.8) to[out=180,in=-135] (x2);
\node at (0.6,-0.6){$A_1$};
  \epic\ee 

We put the Seiberg flipping terms in the last row. Notice that the monopoles change as follows: $\M^{\bullet,0,0}\rightarrow \M^{\bullet,\bullet,0}$ and $\M^{0,0,\bullet}\rightarrow \M^{0,\bullet,\bullet}$.

The quartic term $\tr(b_1 b_2 b_2 b_1)$ became a quadratic term for the $Sp(N-2)$-$Sp(k-2)$ bifundamentals $N_{13}$, which are thus massive and can be integrated out, generating a new quartic term $\tr(b'_1 b'_2 b'_2 b'_1)$. $\hat{A}_1$ is an antisymmetric for $Sp(f-2)$, which we split into an antisymmetric traceless $A_1$ and a trace part $a_1$. Same for $\phi$, antisymmetric traceless for $Sp(N-2)$.

Integrating out the massive fields and removing the $'$'s for notational simplicity, we get theory $3$:
  \be \bpic  \path (-6,0) node[blue](x0) {\small$Sp(\!N\!-\!2\!)$} --    (-3,0) node[blue](x1) {\small$Sp(\!2f\!-\!4\!)$} -- (0,0) node[blue](x2) {\small{$Sp(\!f\!-\!2\!)$}}  -- (-4.5,2.5) node[rectangle,draw](x6) {$\,2f\,$};
  \path (2.5,2.5) node {$ \CW=\g_1 \M^{\bullet,\bullet,\bullet} +\g_2 \M^{\bullet,\bullet,0} + \s_1 \M^{0,\bullet,\bullet}+ \s_2 \M^{0,\bullet,0}  + $}-- (2.5,2) node {$  + M_1 \tr(p b_1  b_1 p) + M_2 \tr(p p) +$}-- (3,1.5) node {$   + \tr(b_1 b b b_1)+ \tr(r p b) +$}-- (3.5,1) node {$ +  \tr(\phi b b) + \tr(A_1 b_1 b_1) + a_1 \tr(b_1 b_1) $};
\draw [-] (x1) to (x0); 
\draw [-] (x1) to (x2); 
\draw [-] (x0) to (x6); 
\draw [-] (x1) to (x6); 
\draw [-] (x6) to[out=45, in=0] (-4.5,3.3) to[out=180,in=135] (x6);
\draw [-] (x6) to[out=45, in=0] (-4.5,3.5) to[out=180,in=135] (x6);
\node at (-3.3,3){$M_1,M_2$};
\draw [-] (x0) to[out=-45, in=0] (-6,-0.8) to[out=180,in=-135] (x0);
\node at (-5.4,-0.6){$\phi$};
\draw [-] (x2) to[out=-45, in=0] (0,-0.8) to[out=180,in=-135] (x2);
\node at (0.6,-0.6){$A_1$};
\node[above right] at (-1.8,0){$b_1$};
\node[above right] at (-4.8,0){$b$};
\node at (-4,1.4){$p$}; 
\node at (-5,1.4){$r$}; 
\node at (-7,2){$\CT_3:$};
  \epic\ee 

where $M_{1,2}, a_1, \g_{1,2}, \s_{1,2}$ are gauge singlets,  $r$ are fundamentals of $Sp(N-2)$, $p$ are fundamentals of $Sp(2(f-2))$.

We can express the $R$-charges of the elementary fields in theory $3$ as a function of the $R$-charges in theory $1$ $r_F$ and $r_\cA$:
\be \ba{l}
r_{b_1} = 1-(r_b)_{\CT_{2}} = r_\cA/2 \\
r_{b} = 1-(r_{\phi_c})_{\CT_{2}}/2 = 1- r_\cA/2 \\
r_{A_1} = r_{a_1} = 2-2r_{b_1} = 2-r_\cA \\
r_{\phi} = 2-2r_{b} =r_{\cA} \\
r_r = (r_p)_{\CT_{2}}+r_\cA/2 = r_F +r_\cA\\
r_p = 1-(r_p)_{\CT_{2}} = 1-(r_\cA/2+r_F) = 1- r_\cA/2 - r_F\\
r_{M_2} = 2-2r_p =2 r_F + r_\cA \\
r_{M_1} = 2-2r_{b_1}-2r_p =  2 r_F
\ea \ee

The mapping of the chiral ring generators of the three theories $\CT_1$, $\CT_2$ and $\CT_3$ is
\be 
\ba{c}\CT_1\\
\tr( Q_i Q_j)  \\
\tr( Q_i \cA Q_j)  \\
\tr( Q_i \cA^J Q_j)  \\
\tr(\cA^J)\\
\tr(\cA^{N-1})\\
\tr(\cA^N)\\
 \{\M_{\cA^J}\}\\
 \{\M_{\cA^{N-2}}\}\\
  \{\M_{\cA^{N-1}}\}
\ea
\qquad \Longleftrightarrow \qquad
\ba{c}\CT_2\\
M_{ij} \\
 \tr(p_i  p_j)\\
 \tr(p_i \phi^{J-1} p_j)\\
\tr(\phi^J)\\
\tr(\phi^{N-1})  \\
 \gamma  \\
 \{\M^{ \bullet,0}_{\phi^J}\} \\
 \{\M^{ \bullet,0}_{\phi^{N-2}}\} \\
 \s
\ea
\qquad \Longleftrightarrow \qquad
\ba{c}\CT_3\\
 (M_1)_{ij}  \\
(M_2)_{ij}  \\
\tr( r_i \phi^{J-2} r_j)  \\
\tr(\phi^J)\\
\g_2\\
\g_1 \\
 \{\M^{ \bullet, 0, 0}_{\phi^J}\}\\
\s_2\\
\s_1  \ea
\quad\qquad
\ba{l}
\,\\
\,\\
\,\\
J=2,\ldots,N-1\\
J=2,\ldots,N-2\\
\, \\
\, \\
J=0,1,\ldots,N-3\\
\, \\
\,  \ea
 \ee

\subsection{After $k$ steps}
After $k$ steps of deconfining and dualizing, we arrive to a quiver with $k+1$ nodes:
  \be \bpic  \path (-6.5,0) node[blue](x0) {\small$Sp(\!N\!-\!k\!)$} --    (-3.5,0) node[blue](x1) {\small$Sp(\!k\!(f\!-\!2)\!)$} -- (0,0) node[blue](x2) {\small$Sp(\!(\!k\!-\!1\!)\!(f\!-\!2)\!)$} -- (3,0) node[blue](x3) {\small{$\ldots$}} -- (6,0) node[blue](x4) {\small{$Sp(\!f\!-\!2\!)$}}; 
  \node at (-5.,2.5)[rectangle,draw](x6) {$\,2f\,$};
  \path (2.5,3.2) node {$ \CW= \sum_{i=1}^{k} \g_i \M^{\bullet^{k-i+2}, 0^{i-1}} +\sum_{i=1}^{k} \s_i \M^{0, \bullet^{k-i+1}, 0^{i-1}}  + $}-- (-1.5,2.5) node[right] {$  +  \sum_{i=1}^{k-1} M_i \tr(p b_{k-1} \ldots b_{i} b_{i}  \ldots b_{k-1} p) +M_k \tr(pp)+$}-- (-1.5,1.8) node[right] {$   + \tr(b_{k-1} b b b_{k-1})+ \tr(r p b)  +  \tr(\phi b b)+$}-- (-1.5,1.1) node[right] {$ + \sum_{i=1}^{k-1} ( \tr(A_i b_i b_i) + \tr(A_i b_{i-1} b_{i-1}) + a_i \tr(b_i b_i)) $};
\draw [-] (x1) to (x0); 
\draw [-] (x1) to (x2); 
\draw [-] (x1) to (x6); 
\draw [-] (x0) to (x6); 
\draw [-] (x2) to (x3); 
\draw [-] (x4) to (x3); 
\draw [-] (x6) to[out=45, in=0] (-5.,3.1) to[out=180,in=135] (x6);
\draw [-] (x6) to[out=45, in=0] (-5.,3.3) to[out=180,in=135] (x6);
\draw [-] (x6) to[out=45, in=0] (-5.,3.7) to[out=180,in=135] (x6);
\node at (-4.2,3.2){$M_i$};
\node at (-5.,3.5){$\ldots$};
\draw [-] (x0) to[out=-45, in=0] (-6.5,-0.8) to[out=180,in=-135] (x0);
\node at (-5.9,-0.6){$\phi$};
\draw [-] (x2) to[out=-45, in=0] (0,-0.8) to[out=180,in=-135] (x2);
\node at (0.8,-0.6){$A_{k-1}$};
\draw [-] (x4) to[out=-45, in=0] (6,-0.8) to[out=180,in=-135] (x4);
\node at (6.6,-0.6){$A_1$};
\node[above right] at (-2.4,-0.1){$b_{k-1}$};
\node[above right] at (-5.2,-0.1){$b$};
\node[above right] at (1.6,-0.1){$b_{k-2}$};
\node[above right] at (4.2,-0.1){$b_{1}$};
\node at (-5.9,1.4){$r$}; 
\node at (-4.6,1.4){$p$}; 
\node at (-6.5,2.3){$\CT_{k+1}:$};
  \epic\ee 

We can express the $R$-charges of the elementary fields in theory $k+1$ as a function of the two independent $R$-charges in theory $1$, $r_F$ and $r_\cA$:
\be \ba{l}
r_{b_i} = r_\cA/2 \qquad i=1,\ldots,k-1 \\
r_{b} =  1- r_\cA/2 \\
r_{A_i} = r_{a_i}  = 2-r_\cA  \qquad i=1,\ldots,k-1\\
r_{\phi} =r_{\cA} \\
r_r = r_F + k \,r_\cA/2\\
r_p = 1- (k-1) r_\cA/2 - r_F\\
r_{M_i} =2 r_F + (i-1) r_\cA  \qquad i=1,\ldots,k
\ea \ee
The mapping of the chiral ring generators with the starting theory $\CT_1$ is
\be 
\ba{c}\CT_1\\
\tr( Q_i \cA^J Q_j)  \\
\tr( Q_i \cA^J Q_j)  \\
\tr(\cA^J)\\
\tr(\cA^J)\\
 \{\M_{\cA^J}\}\\
 \{\M_{\cA^{J}}\} \ea
\qquad \Longleftrightarrow \qquad
\ba{c}\CT_{k+1}\\
 (M_{J+1})_{ij}  \\
\tr( r_i \phi^{J-k} r_j)  \\
\tr(\phi^J)\\
\g_{N-J+1}\\
 \{\M^{ \bullet, 0, 0, \ldots, 0}_{\phi^J}\}\\
\s_{N-J} \ea
\quad\qquad
\ba{l}
\,\\
J=0,\ldots,k-1\\
J=k,\ldots,N-1\\
J=2,\ldots,N-k\\
J=N-k+1,\ldots,N\\
J=0,1,\ldots,N-k-1\\
J=N-k,\ldots,N-1
  \ea
 \ee

\subsection{Fully deconfined tail}
After $N-1$ steps, the leftmost group is $Sp(1)$ so there is no antisymmetric traceless  to deconfine, and we just dualize the $Sp(1)$ using Aharony duality.

 The final result is that our starting theory $\CT_1$
\be \bpic  
\path (-5,0) node{$\CT_1:$} -- (-3,0) node[blue](x1) {$Sp(N)$} -- (-0.5,0) node[rectangle,draw](x2) {\small{$2f$}};
\draw [-] (x1) to (x2); 
\draw [-] (x1) to[out=-45, in=0] (-3,-0.7) to[out=180,in=225] (x1);
\node at (3.5,0){$ \CW= 0$};
\node[below right] at (-3,-0.5){$\cA$};
\node at (-1.6,0.3){$Q_i$}; \epic  \ee
is dual to a fully deconfined quiver with $N$ gauge nodes $\CT_{DEC}$
  \be \label{TDEC} \bpic  \path (-6,0) node[blue](x0) {\small$Sp(\!N(f\!-\!2)\!)$} --    (-2,0) node[blue](x1) {\small$Sp(\!(\!N\!-\!1\!)\!(f\!-\!2)\!)$} -- (1,0) node[blue](x2) {$\ldots$} -- (3,0) node[blue](x3) {\small{$Sp(\!f\!-\!2\!)$}}  -- (-6,2.5) node[rectangle,draw](x6) {$\,2f\,$};
  \path (1,3) node {$ \CW= \sum_{i=1}^{N-1} \g_i \M^{0, \bullet^{N-i}, 0^{i-1}} +\sum_{i=1}^{N} \s_i \M^{\bullet^{N-i+1}, 0^{i-1}}  + $}-- (-3,2.3) node[right] {$  +  \sum_{i=1}^{N} M_i \tr(p b_{N-1} \ldots b_{i} b_{i}  \ldots b_{N-1} p) +$}-- (-3,1.6) node[right]  {$ + \sum_{i=1}^{N-1} ( \tr(A_i b_i b_i) + \tr(A_i b_{i-1} b_{i-1}) + a_i \tr(b_i b_i)) $};
\draw [-] (x1) to (x0); 
\draw [-] (x1) to (x2); 
\draw [-] (x0) to (x6); 
\draw [-] (x2) to (x3); 
\draw [-] (x6) to[out=45, in=0] (-6,3.1) to[out=180,in=135] (x6);
\draw [-] (x6) to[out=45, in=0] (-6,3.3) to[out=180,in=135] (x6);
\draw [-] (x6) to[out=45, in=0] (-6,3.7) to[out=180,in=135] (x6);
\node at (-5.2,3.2){$M_i$};
\node at (-6,3.5){$\ldots$};
\draw [-] (x1) to[out=-45, in=0] (-2,-0.8) to[out=180,in=-135] (x1);
\node at (-1.1,-0.6){$A_{N-1}$};
\draw [-] (x3) to[out=-45, in=0] (3,-0.8) to[out=180,in=-135] (x3);
\node at (3.6,-0.6){$A_1$};
\node[above right] at (-0.5,-0.1){$b_{N-2}$};
\node[above right] at (-4.8,-0.1){$b_{N-1}$};
\node[above right] at (1.5,-0.1){$b_1$};
\node at (-5.8,1.4){$p$}; 
\node at (-7.5,2.3){$\CT_{DEC}:$};
  \epic\ee

The $R$-charges of the elementary fields in the fully deconfined theory are given in terms of the two independent $R$-charges in theory $1$, $r_F$ and $r_\cA$ as follows:
\be \ba{l}
r_{b_i} = r_\cA/2 \qquad i=1,\ldots,N-1 \\
r_{A_i} = r_{a_i}  = 2-r_\cA  \qquad i=1,\ldots,N-1\\
r_p = 1- (N-1) r_\cA/2 - r_F\\
r_{M_i} =2 r_F + (i-1) r_\cA  \qquad i=1,\ldots,N
\ea \ee
The mapping of the chiral ring generators in the duality $\CT_1 \leftrightarrow \CT_{DEC}$ is
\be \label{mapt1tdec}
\ba{c}\CT_1 \\
\tr( Q_i \cA^J Q_j)  \\
\tr(\cA^J)\\
 \{\M_{\cA^{J}}\} \ea
\qquad \Longleftrightarrow \qquad
\ba{c} \CT_{DEC} \\
 (M_{J+1})_{ij}  \\
\g_{N-J+1}\\
\s_{N-J} \ea
\quad\qquad
\ba{l}
\,\\
J=0,\ldots,N-1\\
J=2,\ldots,N\\
J=0,\ldots,N-1
  \ea
 \ee

Notice that all chiral ring generators of $\CT_1$ map to gauge singlets in $\CT_{DEC}$. This is similar to Aharony duality for $Sp$ gauge group without rank-$2$ matter fields. Also, in the case of $Sp(1)$ gauge group, our duality $\CT_1 \leftrightarrow \CT_{DEC}$ reduces to Aharony duality.


\subsection{Superpotential deformation: $\CW= \tr(Q_{2f-1} \cA^J Q_{2f})$}
In this section and in section \ref{sec:N=4}, we discuss complex deformations of the duality between our original theory with a single node $Sp(N)$ \eqref{T1} $\CT_1$ and the fully deconfined quiver $\CT_{DEC}$ \eqref{TDEC}. As usual in Seiberg like dualities, a complex deformation on the electric side will induce a Higgsing of the gauge groups on the magnetic side.

In this section we consider a superpotential deformation in $\CT_1$ of the form $\tr(Q_{2f-1} \cA^J Q_{2f})$. This meson, according to \eqref{mapt1tdec}, is mapped in $\CT_{DEC}$ to  $(M_{J+1})_{2f-1, f}$, the flipping field for the meson $\tr(p_{2f-1} \ldots b_{J+1} b_{J+1}  \ldots p_{2f})$. We take $J<N$. Turning on a linear superpotential term in $M_J$ means that this $2f \times 2f$ matrix of long mesons must take a non-zero vacuum expectation value of minimal non zero rank. This is achieved by giving a vev to the bifundamentals $b_J, b_{J+1}, \ldots, b_{N-1}$ and to the flavors $p$, such that $\tr(p_{2f-1} b_{N-1} \ldots b_{J+1} b_{J+1}  \ldots b_{N-1} p_{2f})=1$, while all other mesons are zero. 

 Without going too much into the details, the final result is that in \eqref{TDEC} the $N-J$ gauge groups on the left
 $$Sp(N(f-2)),\ldots, Sp(h(f-2)), \ldots, Sp((J+1)(f-2))$$ 
 are Higgsed to 
 $$Sp(N(f-2)-(N-J)),\ldots, Sp(h(f-2)-(h-J)), \ldots, Sp((J+1)(f-2)-(J+1-J))\,,$$
while the $J$ remaining gauge groups on the right are not Higgsed. The last flavor $p_{2f-1},p_{2f}$ \emph{migrates} from the left-most node to node $Sp(J(f-2))$. More precisely, since the node $Sp((J+1)(f-2))$ is Higgsed down to $Sp((J+1)(f-3)+J)$, the bifundamental field $b_{J}$ splits into a new $Sp((J+1)(f-3)+J) - Sp(J(f-2))$ bifundamental and a flavor for the node $Sp(J(f-2))$. The flipping fields for the mesons split into two sets $M_H$, $H=1,\ldots,N$, and $(M')_K$, $K=1,\ldots,J$.

 The final result is that 
\be \bpic  
\path (-3,0) node[blue](x1) {$Sp(N)$} -- (-0.5,0) node[rectangle,draw](x2) {\small{$2f$}};
\draw [-] (x1) to (x2); 
\draw [-] (x1) to[out=-45, in=0] (-3,-0.7) to[out=180,in=225] (x1);
\node at (3.5,0){$ \CW= \tr(Q_{f-1} \cA^J Q_f)$};
\node[below right] at (-3,-0.5){$\cA$};
\node at (-1.6,0.3){$Q_i$}; \epic  \ee
is dual to
  \be \label{TDEC1} \bpic  \path (-6,0) node[blue](x0) {\small$\!Sp\!(\!N\!(\!f\!-\!3\!)\!+\!J\!)\!$} --    (-2.2,0) node[blue](x1) {\small$\!Sp\!(\!(\!N\!-\!1\!)\!(\!f\!-\!3\!)\!+\!J\!)\!$} -- (0.8,0) node[blue](x2) {$\ldots$} -- (3,0) node[blue](x21) {\small$Sp(\!J\!(f\!-\!2)\!)$} -- (5.3,0) node[blue](x22) {$\ldots$} -- (7,0) node[blue](x3) {\small{$Sp(\!f\!-\!2\!)$}}  -- (-6,1.75) node[rectangle,draw](x6) {$\!2f\!-\!2\!$} -- (3,1.75) node[rectangle,draw](x7) {$\,2\,$};
  \path (-1,4) node {$ \CW= \sum_{i=1}^{N-1} \g_i \M^{0, \bullet^{N-i}, 0^{i-1}} +\sum_{i=1}^{N} \s_i \M^{\bullet^{N-i+1}, 0^{i-1}}  + $}--
   (-5,2.6) node[right] {$  +  \sum_{i=1}^{N} M_i \tr(p b_{N-1} \ldots b_{i} b_{i}  \ldots b_{N-1} p) + $}--
   (-5,1.9) node[right] {$ + \sum_{i=1}^{J} M'_i \tr(p' b_{J-1} \ldots b_{i} b_{i}  \ldots b_{J-1} p') $} --
    (-5,3.3) node[right]  {$ + \sum_{i=1}^{N-1} ( \tr(A_i b_i b_i) + \tr(A_i b_{i-1} b_{i-1}) + a_i \tr(b_i b_i)) +$};
\draw [-] (x1) to (x0); 
\draw [-] (x1) to (x2); 
\draw [-] (x0) to (x6); 
\draw [-] (x2) to (x21); 
\draw [-] (x21) to (x22); 
\draw [-] (x22) to (x3); \draw [-] (x21) to (x7); 
\draw [-] (x6) to[out=45, in=0] (-6,2.35) to[out=180,in=135] (x6);
\draw [-] (x6) to[out=45, in=0] (-6,2.55) to[out=180,in=135] (x6);
\draw [-] (x6) to[out=45, in=0] (-6,2.95) to[out=180,in=135] (x6);
\node at (-6.8,2.3){$M_i$};
\node at (-6,2.75){$\ldots$};
\draw [-] (x7) to[out=45, in=0] (3,2.3) to[out=180,in=135] (x7);
\draw [-] (x7) to[out=45, in=0] (3,2.5) to[out=180,in=135] (x7);
\draw [-] (x7) to[out=45, in=0] (3,2.8) to[out=180,in=135] (x7);
\node at (3.8,2.3){$M'_i$};
\node at (3,2.6){$\ldots$};
\draw [-] (x1) to[out=-45, in=0] (-2.2,-0.8) to[out=180,in=-135] (x1);\node at (-1.3,-0.6){$A_{N-1}$};
\draw [-] (x21) to[out=-45, in=0] (3,-0.8) to[out=180,in=-135] (x21);\node at (3.6,-0.6){$A_{J}$};
\draw [-] (x3) to[out=-45, in=0] (7,-0.8) to[out=180,in=-135] (x3);\node at (7.6,-0.6){$A_1$};
\node[above right] at (-4.9,-0.1){$b_{N-1}$};
\node[above right] at (-0.6,-0.1){$b_{N-2}$};
\node[above right] at (1.3,-0.1){$b_{J}$};
\node[above right] at (3.9,-0.1){$b_{J-1}$};
\node[above right] at (5.6,-0.1){$b_1$};
\node at (-5.8,0.9){$p$}; \node at (2.8,0.9){$p'$}; 
  \epic\ee 

\paragraph{Complex mass deformation}
  Let us start from $f>2$ and turn on a complex mass for $2$ flavors, $\delta \CW= \tr(Q_{2f-1} Q_{2f})$. This is the special case $J=0$ of the discussion above. 
The flavor $p'$ and the gauge singlets $M'$ are absent for $J=0$, and all the gauge groups in $\CT_{DEC}$ get partially Higgsed. The final result is precisely \eqref{TDEC}  with $f \rightarrow f-1$. We thus get a consistency check of the duality $\CT_1 \leftrightarrow \CT_{DEC}$.

\subsection{\emph{$\CN\!=\!4$-like} deformation: $ \CW=\sum_{j} \tr(Q_{2j-1} \cA Q_{2j})$}\label{sec:N=4}
We now consider adding $f$ cubic terms to $\CT_1$, obtaining $Sp(N)$ with $2f$ chiral flavors and $ \CW=\sum_{j=1}^f \tr(Q_{2j-1} \cA Q_{2j})$. Using the results just obtained in \eqref{TDEC1}, on the dual side, all the flavors \emph{migrate} to the right-most node $Sp(f-2)$, and out the tower of singlets $(M_J)_{ij}$, only $(M_1)_{ij}$ survive. The tail of gauge groups 
$$Sp(N(f-2)),\ldots,Sp(3(f-2)), Sp(2(f-2)), Sp(f-2)$$ 
Higgs to 
$$Sp(f-2N),\ldots,Sp(f-6), Sp(f-4), Sp(f-2)\,.$$ 
Notice that the right-most gauge group $Sp(f-2)$ is not Higgsed.

The final result is that 
\be \bpic  
\path (-5,0) node{$\CT_1:$} -- (-3,0) node[blue](x1) {$Sp(N)$} -- (-0.5,0) node[rectangle,draw](x2) {\small{$2f$}};
\draw [-] (x1) to (x2); 
\draw [-] (x1) to[out=-45, in=0] (-3,-0.7) to[out=180,in=225] (x1);
\node at (3.5,0){$ \CW= \sum_{j=1}^f \tr(Q_{2j-1} \cA Q_{2j})$};
\node[below right] at (-3,-0.5){$\cA$};
\node at (-1.9,0.3){$Q_i$}; \epic  \ee
is dual to\footnote{The result \eqref{TDEC4} can also be obtained deforming the duality $\CT_1 \leftrightarrow \CT_2$ discussed in section \ref{step1Sp}, recalling that $\tr(Q \cA Q) \leftrightarrow \tr(pp)$: all the fundamentals $p$ in $\CT_2$ becomes massive.  At this point one sequentially deconfines the antisymmetric, building the tail without carrying around the flavors, which remain attached to the right-most node $Sp(f-2)$.}
  \be \label{TDEC4} \bpic  \path (-6,0) node[blue](x0) {\small$Sp(\!f\!-\!2N\!)$} --    (-2,0) node[blue](x1) {\small$Sp(\!f\!-\!2N\!+\!2\!)\!$} -- (1,0) node[blue](x2) {$\ldots$} -- (3,0) node[blue](x3) {\small{$Sp(\!f\!-\!2\!)$}}  -- (3,2.5) node[rectangle,draw](x6) {$\,2f\,$};
  \path (-3,3) node {$ \CW= \sum_{i=1}^{N-1} \g_i \M^{0, \bullet^{N-i}, 0^{i-1}} +\sum_{i=1}^{N} \s_i \M^{\bullet^{N-i+1}, 0^{i-1}}  + $}-- (-3,2.3) node {$  + M \tr(p p) +$}-- (-3,1.6) node  {$ + \sum_{i=1}^{N-1} ( \tr(A_i b_i b_i) + \tr(A_i b_{i-1} b_{i-1}) + a_i \tr(b_i b_i)) $};
\draw [-] (x1) to (x0); 
\draw [-] (x1) to (x2); 
\draw [-] (x3) to (x6); 
\draw [-] (x2) to (x3); 
\draw [-] (x6) to[out=45, in=0] (3,3.3) to[out=180,in=135] (x6);
\node at (3.7,3.2){$M$};
\draw [-] (x1) to[out=-45, in=0] (-2,-0.8) to[out=180,in=-135] (x1);
\node at (-1.1,-0.6){$A_{N-1}$};
\draw [-] (x3) to[out=-45, in=0] (3,-0.8) to[out=180,in=-135] (x3);
\node at (3.6,-0.6){$A_1$};
\node[above right] at (-0.5,-0.1){$b_{N-2}$};
\node[above right] at (-4.8,-0.1){$b_{N-1}$};
\node[above right] at (1.5,-0.1){$b_1$};
\node at (3.2,1.4){$p$}; 
  \epic\ee 
This result is strictly speaking valid for $f>2N$. If $f \leq 2N$ the dual quiver becomes shorter and some of the flipping fields $\g$ and $\s$ decouple. This is due to the fact the theory $Sp(N)$ with $\CW= \sum_{j=1}^f \tr(Q_{2j-1} \cA Q_{2j})$ if $f \leq 2N$ becomes ``bad'' in the Gaiotto-Witten sense, so some Coulomb branch operators (that is $\tr(\CA^h)$ and $\{\M_{A^k}\}$) become free and decouple.
   
\subsection{Real masses and Chern-Simons terms}\label{addCS}
Starting from the dualities discussed above, it is easy to turn a Chern-Simon interaction at level $k$: we simply start from the theory with $2f+2k$ flavors and turn a positive real mass for $2k$ flavors. We obtain $Sp(N)_k$ with $2f$ flavors and $\CW=0$. Now $f$ and $k$ are either integers or half-integers, but $f+k$ is always an integer. 

The real mass is in the supermultiplet of the $U(2f+2k)$ global symmetry current, so in the fully deconfined dual \eqref{TDEC} the real mass is  mapped to a real mass for some of the flavors $p$ (the bifundamental fields $b_i$ are not charged under the $U(2f+2k)$ global symmetry) and some of the gauge singlets $M$. In the fully deconfined dual \eqref{TDEC} (with $f \rightarrow f+k$), $2k$ flavors $p$'s get a negative real mass, which induces a negative Chern-Simons level $-k$ for the leftmost node, while the Chern-Simons levels of all the other nodes do not get any shift.

If $k\neq 0$, the monopoles $\{\M_{\cA^J}\}$ are not in the chiral ring of $Sp(N)_k$, accordingly the singlet fields $\s_i$ disappear from the deconfined dual of $Sp(N)_k$. 

Summing up, the fully deconfined dual of $Sp(N)_k$ with antisymmetric and $2f$ flavors, $\CW=0$, is
\footnote{\label{FNlowr}If $f+|k|=2$, $k=0,\pm 1, \pm 2$, all the ranks in the quiver tails vanish. In the case of the fully deconfined theory, the full gauge group is trivial. This means that the deconfined theory is replaced by a Wess-Zumino, with a non trivial superpotential constructed out of the gauge singlet fields $\g_i, \s_i, M_i$, as in  \cite{Amariti:2018wht, Benvenuti:2018bav}. 

If $f+|k|=3$, $k=0,\pm 1, \pm 2, \pm 3$, the quiver tail is $Sp(N)_{-k} - Sp(N-1)- Sp(N-2) -\ldots - Sp(1)$, and this can be \emph{sequentially confined}. We start from the right-most node, which is a $Sp(1)_0$ with $2\cdot2$ flavors, so it confines. Moreover, the antisymmetric for the $Sp(2)_0$ node is removed. We then dualize the $Sp(2)_0$ with $3\cdot 2$ flavors, which also confines. After $N-1$ confining steps, we end up with $Sp(N)_{-k}$, with an antisymmetric plus $6-2|k|$ flavors, and some flipping fields. The same $Sp(N)_{+k} \leftrightarrow Sp(N)_{-k}$ duality can be achieved in a different way, see \cite{Amariti:2018wht} and eq. 5.2 in \cite{Benvenuti:2018bav}.}

  \be\label{TDECk} \bpic  \path (-6,0) node[blue](x0) {\small$Sp(\!N\!(\!f\!+\!k\!-\!2\!)\!)_{-k}$} --    (-2,0) node[blue](x1) {\small$Sp(\!(\!N\!-\!1\!)\!(\!f\!+\!k\!-\!2\!)\!)_0$} -- (1,0) node[blue](x2) {$\ldots$} -- (3,0) node[blue](x3) {\small{$Sp(\!f\!+\!k\!-\!2\!)_0$}}  -- (-6,2.5) node[rectangle,draw](x6) {$\,2f\,$};
  \path (1,3) node {$ \CW= \sum_{i=1}^{N-1} \g_i \M^{0, \bullet^{N-i}, 0^{i-1}}  + $}-- (-3,2.3) node[right] {$  +  \sum_{i=1}^{N} M_i \tr(p b_{N-1} \ldots b_{i} b_{i}  \ldots b_{N-1} p) +$}-- (-3,1.6) node[right]  {$ + \sum_{i=1}^{N-1} ( \tr(A_i b_i b_i) + \tr(A_i b_{i-1} b_{i-1}) + a_i \tr(b_i b_i)) $};
\draw [-] (x1) to (x0); 
\draw [-] (x1) to (x2); 
\draw [-] (x0) to (x6); 
\draw [-] (x2) to (x3); 
\draw [-] (x6) to[out=45, in=0] (-6,3.1) to[out=180,in=135] (x6);
\draw [-] (x6) to[out=45, in=0] (-6,3.3) to[out=180,in=135] (x6);
\draw [-] (x6) to[out=45, in=0] (-6,3.7) to[out=180,in=135] (x6);
\node at (-5.2,3.2){$M_i$};
\node at (-6,3.5){$\ldots$};
\draw [-] (x1) to[out=-45, in=0] (-2,-0.8) to[out=180,in=-135] (x1);
\node at (-1.1,-0.6){$A_{N-1}$};
\draw [-] (x3) to[out=-45, in=0] (3,-0.8) to[out=180,in=-135] (x3);
\node at (3.6,-0.6){$A_1$};
\node[above right] at (-0.5,-0.1){$b_{N-2}$};
\node[above right] at (-4.8,-0.1){$b_{N-1}$};
\node[above right] at (1.5,-0.1){$b_1$};
\node at (-5.8,1.4){$p$}; 
  \epic\ee 
  
The relation among $R$-charges of the elementary fields is the same as before:
\be \ba{l}
r_{b_i} = r_\cA/2 \qquad i=1,\ldots,N-1 \\
r_{A_i} = r_{a_i}  = 2-r_\cA  \qquad i=1,\ldots,N-1\\
r_p = 1- (N-1) r_\cA/2 - r_F\\
r_{M_i} =2 r_F + (i-1) r_\cA  \qquad i=1,\ldots,N
\ea \ee
The mapping of the chiral ring generators  is
\be 
\ba{c}Sp(N)_k\,,\,\,\CW\!=\!0 \\
\tr( Q_i \cA^J Q_j)  \\
\tr(\cA^J)\ea
\qquad \Longleftrightarrow \qquad
\ba{c}\textrm{Theory}\,\, \eqref{TDECk}\\
 (M_{J+1})_{ij}  \\
\g_{N-J+1} \ea
\quad\qquad
\ba{l}
\,\\
J=0,\ldots,N-1\\
J=2,\ldots,N
  \ea
 \ee

\subsubsection*{Deconfining with non zero Chern-Simons interactions}
It is instructive to see how to reach the result \eqref{TDECk} deconfining and dualizing sequentially the $Sp(N)_k$ theory, as done before in the case of vanishing Chern-Simons coefficient. The procedure is pretty much the same, difference is that with non zero that Chern-Simons the relevant duality is \cite{Willett:2011gp, Benini:2011mf}
\be\label{WY} \ba{c} Sp(N)_k \,\, \textrm{w/} \,\, 2F \,\, \textrm{flavors},\\ \CW=0 \ea
 \Longleftrightarrow 
 \ba{c}Sp(F+|k|-N-1)_{-k}\,\, \textrm{w/} \,\, 2F \,\, \textrm{flavors}\,\, p_i,\\ \CW= A_1^{ij} \tr(p_i p_j) \ea \ee

It is important that the duality \eqref{WY} generates a Chern-Simons interaction for the global $U(2F)$ symmetry, with level $+k$ \cite{Benini:2011mf}. In our case the global $U(2F)$ symmetry is partially gauged. Such Chern-Simons interaction has the effect that when we dualize a $Sp$ node in a quiver, the Chern-Simons level of the nearby nodes in the quiver is shifted by $+k$. After $h$ steps of deconfining and dualizing, one reaches the partially deconfined theory:

  \be \label{TDECcs} \bpic  \path (-6,0) node[blue](x0) {\small$Sp(\!N\!-\!h\!)_k$} --    (-3,0) node[blue](x1) {\small$Sp(\!h\!(\!f\!+\!k\!-\!2\!)\!)_{-k}$} -- (1,0) node[blue](x2) {\small$Sp(\!(h\!-\!1)\!(\!f\!+\!k\!-\!2\!)\!)_0$} -- (4,0) node[blue](x3) {\small{$\ldots$}} -- (7,0) node[blue](x4) {\small{$Sp(\!f\!+\!k\!-\!2\!)_0$}}; 
  \node at (-4.5,2.5)[rectangle,draw](x6) {$\,2f\,$};
  \path (2.5,3.2) node {$ \CW= \sum_{i=1}^{h} \g_i \M^{\bullet^{h-i+2}, 0^{i-1}}  + $}-- (-1.5,2.5) node[right] {$  +  \sum_{i=1}^{h-1} M_i \tr(p b_{h-1} \ldots b_{i} b_{i}  \ldots b_{h-1} p) +M_h \tr(pp)+$}-- (-1.5,1.8) node[right] {$   + \tr(b_{h-1} b b b_{h-1})+ \tr(r p b)  +  \tr(\phi b b)+$}-- (-1.5,1.1) node[right] {$ + \sum_{i=1}^{h-1} ( \tr(A_i b_i b_i) + \tr(A_i b_{i-1} b_{i-1}) + a_i \tr(b_i b_i)) $};
\draw [-] (x1) to (x0); 
\draw [-] (x1) to (x2); 
\draw [-] (x1) to (x6); 
\draw [-] (x0) to (x6); 
\draw [-] (x2) to (x3); 
\draw [-] (x4) to (x3); 
\draw [-] (x6) to[out=45, in=0] (-4.5,3.1) to[out=180,in=135] (x6);
\draw [-] (x6) to[out=45, in=0] (-4.5,3.3) to[out=180,in=135] (x6);
\draw [-] (x6) to[out=45, in=0] (-4.5,3.7) to[out=180,in=135] (x6);
\node at (-3.7,3.2){$M_i$};
\node at (-4.5,3.5){$\ldots$};
\draw [-] (x0) to[out=-45, in=0] (-6,-0.8) to[out=180,in=-135] (x0);
\node at (-5.4,-0.6){$\phi$};
\draw [-] (x2) to[out=-45, in=0] (1,-0.8) to[out=180,in=-135] (x2);
\node at (1.8,-0.6){$A_{h-1}$};
\draw [-] (x4) to[out=-45, in=0] (7,-0.8) to[out=180,in=-135] (x4);
\node at (7.6,-0.6){$A_1$};
\node[above right] at (-1.7,-0.1){$b_{h-1}$};
\node[above right] at (-5,-0.1){$b$};
\node[above right] at (2.5,-0.1){$b_{h-2}$};
\node[above right] at (4.8,-0.1){$b_{1}$};
\node at (-5.6,1.4){$r$}; 
\node at (-4,1.4){$p$}; 
\node at (-6.5,2.3){$\CT_{h+1}:$};
  \epic\ee

\section{A sequence of duals for $U(N)$ with an adjoint}\label{adjU}
In this section we find dual descriptions of $U(N)$ with a field $\Phi$ in the adjoint representation and $(F,F)$ flavors $Q_i,\Qt_i$, $\cW=0$. The adjoint field is a $SU(N)$-adjoint, that is $\Phi$ is traceless.

We consider $F \geq 2$: if $F=1$, following the same procedure of deconfining and dualizing, one gets a fully deconfined dual which is a Wess-Zumino model, see \cite{Pasquetti:2019uop}.

The procedure is very similar to the one described in section \ref{asymmSp}, so in this section we give a bit less detail. There are $2N$ dual theories, that are quivers with a number of nodes ranging from $1$ to $N$. The fully deconfined dual has $N$ nodes. 

The main difference with respect to the case of \ref{asymmSp} is that we deconfine the adjoint using the ``one monopole duality" of \cite{Benini:2017dud}, which introduce superpotential terms in the quiver which are linear in the monopoles. Such terms break the topological symmetries and give rise to some complications, for instance the $R$-charge of the monopoles $\M^{\ldots,+,\ldots}$ is not equal to the $R$-charge of the monopoles $\M^{\ldots,-,\ldots}$. (In linear quivers made of $U$ gauge groups, we denote by $\M^{0,0,\pm,\pm,\ldots}$ monopoles with non-zero minimal flux in the nodes with $\pm$ and zero flux in nodes with $0$). In detail, the presence of a linear monopole superpotential leads to a modification of the usual $R$-charge monopole formula; in fact, every time we have a superpotential term $\mathcal W=\M^{\ldots,-,\ldots}$ we need to ensure the marginality of such monopole. The main idea is to start with a simple ansatz for the additional corrections to the standard monopole $R$-charge formula, and fix the additional terms using the marginality of the monopoles contained in the superpotential and the operator map across duality to completely fix the coefficients of such terms. 
Physically, the added terms corresponds to mixed contact terms  between $R$-symmetry and gauge symmetry, that may be computed, for instance, using localisation techniques. However, this goes beyond the aim of the present work.  

Let us now explain a bit more in detail the procedure we are going to use. As we said, monopole operators in the superpotential are not symmetric under charge conjugation. Thus, the modification of the usual $R$-charge formula should distinguish the different signs of the fluxes, so, given a general linear quiver with $N$ gauge nodes, we start from the ansatz:
\be\label{modRch}
	R[\mathfrak M^{m^{(1)}, m^{(2)}, \cdots, m^{(N)}}]=(\text{standard}) \,+ \,\alpha_1 \sum_{i_1=1}^{N_1} m^{(1)}_{i_1}+\cdots+ \,\alpha_N \sum_{i_n=1}^{N_n} m^{(N)}_{i_n},
\ee
where $(\text{standard})$ refers to the usual $R$-charge contributions from matter fields and gauginos, for instance for the following quiver with matter in the (bi-)fundamental and adjoint
\be
\begin{tikzpicture}[baseline]
\scalebox{1.1}{
\tikzstyle{every node}=[font=\footnotesize]
\node[draw=none] (g1) at (0,0) {$U(N_1)$};
\node[draw=none] (g2) at (2.5,0) {$U(N_2)$};
\node[draw=none] (g3) at (5,0) {$\cdots$};
\node[draw=none] (g4) at (7.5,0) {$U(N_n)$};
\node[draw, rectangle] (f1) at (10,0) {$F$};
\draw[transform canvas={yshift=2.5pt},->] (g1) to (g2) ;
\draw[transform canvas={yshift=-2.5pt},<-] (g1) to (g2) ;
\node[draw=none] at (1.3, -0.4) {$B_1, \Bt_1$} ;
\draw[black,solid] (g1) edge [out=45,in=135,loop,looseness=4] (g1);
\node[draw=none] at (0.06, 1.1) {$\Phi_1$} ;
\draw[transform canvas={yshift=2.5pt},->] (g2) to (g3) ;
\draw[transform canvas={yshift=-2.5pt},<-] (g2) to (g3) ;
\node[draw=none] at (3.8, -0.4) {$B_2, \Bt_2$} ;
\draw[black,solid] (g2) edge [out=45,in=135,loop,looseness=4] (g2);
\node[draw=none] at (2.55, 1.1) {$\Phi_2$} ;
\draw[transform canvas={yshift=2.5pt},->] (g3) to (g4) ;
\draw[transform canvas={yshift=-2.5pt},<-] (g3) to (g4) ;
\node[draw=none] at (6.3, -0.4) {$B_{n-1}, \Bt_{n-1}$} ;
\draw[black,solid] (g4) edge [out=45,in=135,loop,looseness=4] (g4);
\node[draw=none] at (7.55, 1.1) {$\Phi_n$} ;
\draw[transform canvas={yshift=2.5pt},->] (g4) to (f1) ;
\draw[transform canvas={yshift=-2.5pt},<-] (g4) to (f1) ;
\node[draw=none] at (9, -0.4) {$Q, \Qt$} ;
}
\end{tikzpicture}
\ee
it reads
\be
\begin{split}
	R[\mathfrak M^{m^{(1)}, m^{(2)}, \cdots, m^{(n)}}]&=(1-r_{\Phi_1})\sum_{i_1 < i_2}^{N_1} |m^{(1)}_{i_1} - m^{(1)}_{i_2}| 
	+\dots +(1-r_{\Phi_n}) \sum_{i_1 < i_2}^{N_n} |m^{(n)}_{i_1} - m^{(n)}_{i_2}| + \\
	&+(1-r_{B_1}) \sum_{i=1}^{N_1} \sum_{j=1}^{N_2} |m^{(1)}_{i} - m^{(2)}_{j}| + 
	\cdots +(1-r_{B_{n-1}}) \sum_{i=1}^{N_{n-1}} \sum_{j=1}^{N_n} |m^{(n-1)}_{i} - m^{(n)}_{j}| + \\
	& + F(1-r_Q) \sum_{i=1}^{N_n} |m^{(n)}_i| -  \sum_{i_1 < i_2}^{N_1} |m^{(1)}_{i_1} - m^{(1)}_{i_2}|- \dots - \sum_{i_1 < i_2}^{N_n} |m^{(n)}_{i_1} - m^{(n)}_{i_2}|
\end{split}
\ee
The parameters $\alpha_i$ are the ones that will be fixed imposing the marginality of the monopoles in the superpotential and the use of the duality map. The use of the duality map can be considered as a weakness of such an effective procedure: given a general quiver theory with an arbitrary combinations of linear monopole superpotential we are not able to provide an expression for the monopole $R$-charge; moreover, in this way we may only find the parameters $\alpha_i$ only in terms of the mixing parameters of the the starting theory. Nonetheless, as we will concretely see later, the procedure we employ works perfectly in order to study the deconfinement of a traceless $U(N)$ adjoint field. 


A first check of the validity of the procedure is that the result for the parameter fixed via the operator map does not depend on which operator we map. Another strong test comes from the computation of the supersymmetric index, where the presence of such contact terms is crucial, since it enters the sum over the gauge magnetic fluxes. 


We start from the case of vanishing Chern-Simons interactions, with this result, it will be easy to turn on a real mass deformation and hence a Chern-Simons term in section \ref{sec:addCSU}, where we discuss the duals $U(N)_k$ with adjoint and flavors.\\

We start with theory $\CT_1$, that is $U(N)$ with a traceless antisymmetric field $\Phi$ and $F$  flavors $Q_i,\Qt_i$. We take the  superpotential to be vanishing, $\cW=0$. Using the standard quiver notation for theories with four supercharges, $\CT_1$ reads
\be
\label{T1U}
\begin{tikzpicture}[baseline]
\scalebox{1.1}{
\tikzstyle{every node}=[font=\footnotesize]
\node[draw=none]at (-2,0) {$\CT_1:$};
\node[draw=none] (g1) at (0,0) {$U(N)$};
\node[draw, rectangle] (f1) at (2.5,0) {$F$};
\draw[transform canvas={yshift=2.5pt},->] (g1) to (f1) ;
\draw[transform canvas={yshift=-2.5pt},<-] (g1) to (f1) ;
\node[draw=none] at (1.3, -0.5) {$Q, \Qt$} ;
\draw[black,solid] (g1) edge [out=45,in=135,loop,looseness=4] (g1);
\node[draw=none] at (0, 1.2) {$\Phi$} ;
\node[draw=none] at (4.5, 0) {$\CW=0$} ;
}
\end{tikzpicture}
\ee

Throughout most of this section, the square node denotes a  $SU(F) \times SU(F)$ global symmetry. 

The global symmetry is $SU(F)^2 \times U(1)_{Q} \times U(1)_{\Phi} \times U(1)_{\text{topological}}$.

The chiral ring is generated by the (dressed) mesons $\tr(\Qt_i \Phi^l Q_j)$, $l=0,\ldots,N-1$, the powers of the antisymmetric traceless field $\tr(\Phi^j)$, $j=2,\ldots,N$, and  the (dressed) monopoles $\{\M_{\Phi^k}\}$, $k=0,1,\ldots,N-1$. In terms of the $R$-charges of the elementary fields $Q_i$ and $\Phi$, which we denote $r_F$ and $r_\Phi$, the $R$-charge of the basic, undressed, monopole $\M$ is
\be R[\M^{\pm}]_{\CT_{1}} = F(1-r_F) + (N-1)(-r_\Phi) \ee

\subsection{Deconfine and dualize with the one-monopole duality}\label{step1Sp}
In order to deconfine the adjoint field, we use a variation of the \emph{confining} one monopole duality of \cite{Benini:2017dud}, which reads
\be\label{decadjU} \ba{c} U(N-1) \, \text{w/ $(N,N)$ flavors}\, q_i,\qt_i \\ \cW= \M^- \ea
           \Llra
\ba{c}  \text{Wess-Zumino with}\,\, N^2+1 \,\, \text{ chirals}\,\, \Phi,s\\ \cW= s \, \det(\Phi)    \ea \ee
 In this duality $\qt q \leftrightarrow \Phi$ and $\M^+ \leftrightarrow s$. 
 
 The mapping $\M^+ \leftrightarrow s$ is in agreement with the $R$-charge computation. On the l.h.s. the topological symmetry is broken by the superpotential term, so the $R$-charge of the monopoles mixes with the topological symmetry:
 \be R[\M^\pm]= (N-1) (1-r_q)-(N-2) \pm \delta \ee
Imposing $R[\M^-]=2$ we get $\delta=-(N-1)r_q$ and thus 
\be R[\M^+]=2-2Nr_q \,.\ee
 On the r.h.s. $R[s]=2- N R[\Phi] = 2-2Nr_q$.

We will need the following  variation of \eqref{decadjU}: we start from \eqref{decadjU}, flip the monopole $\M^+$ in the l.h.s. with a gauge singlet $\g$, on the r.h.s. a superpotential term $s \g$ arises, $s$ and $\g$ become massive, integrating them out the superpotential becomes zero and we obtain the following deconfining duality: 
\be  \label{decU}
 \ba{c} U(N-1) \, \text{w/ $(N,N)$ flavors}\, q_i,\qt_i \\ \cW= \M^- + \g \, \M^{+} \ea
           \Llra
\ba{c}  N^2 \,\, \text{free chirals } \, M^i_j \\ \text{bifundamental of}\,\,SU(N)^2    \ea \ee 
In this duality the chiral ring generators are only the quadratic mesons, which map as $\tr(q^i \qt_j) \leftrightarrow M^i_j$.

Starting from theory $\CT_{1}$, we use \eqref{decU} to \emph{deconfine the adjoint field} into a two-node quiver theory, that is we consider theory $\CT_{1'}$:
\be
\begin{tikzpicture}[baseline]
\scalebox{1.1}{
\tikzstyle{every node}=[font=\footnotesize]
\node[draw=none]at (-2,0) {$\CT_{1'}:$};
\node[draw=none] (g1) at (0,0) {$U(N-1)$};
\node[draw=none] (g2) at (2.5,0) {$U(N)$};
\node[draw, rectangle] (f1) at (5,0) {$F$};
\draw[transform canvas={yshift=2.5pt},->] (g1) to (g2) ;
\draw[transform canvas={yshift=-2.5pt},<-] (g1) to (g2) ;
\node[draw=none] at (1.4, 0.5) {$b', \bt'$} ;
\draw[transform canvas={yshift=2.5pt},->] (g2) to (f1) ;
\draw[transform canvas={yshift=-2.5pt},<-] (g2) to (f1) ;
\node[draw=none] at (3.9, 0.5) {$Q, \Qt$} ;
\node[draw=none] at (2.4, -0.8) {$\CW= \M^{-,0} + \gamma \, \M^{+,0}+ \b \tr(b' \bt')$} ;
}
\end{tikzpicture}
\ee
%
%

In $\CT_{1}$ the monopoles $\M^{0,\pm}$, $\M^{+,+}$ and $\M^{--}$ are non trivial elements of the chiral ring, their $R$-charges read
\be \ba{rcl} 
R[\M^{0,+}]_{\CT_{1'}} &=& F(1-r_Q)  \\
R[\M^{0,-}]_{\CT_{1'}} &=& F(1-r_Q) - 2(N-1) r_{b'} \\
 R[\M^{+,+}]_{\CT_{1'}} &=& F(1-r_Q)  - 2(N-1) r_{b'} \\
 R[\M^{-,-}]_{\CT_{1'}} &=& F(1-r_Q)  - 2(N-2) r_{b'} 
  \ea\ee
 
Using that $r_Q=r_F, \, r_\Phi=2 r_{b'}$ and $R[\M^{\pm}]_{\CT_{1}} = F(1-r_F) - (N-1) r_\Phi$, we see that these monopoles map into $\CT_1$ as follows
\be 
\ba{c} \CT_{1'}\\
 \M^{ 0,+}\\
 \M^{ 0,-}\\
 \M^{+,+}\\
  \M^{-,-}
\ea
\qquad \Longleftrightarrow \qquad
\ba{c}\CT_1 \\
 \{\M^+_{\Phi^{N-1}}\}\\
 \M^{-} \\
 \M^{+} \\
  \{\M^{-}_{\Phi}\}
\ea
 \ee
 
From the mapping we learn the following rule: deconfining and adjoint with the one monopole duality  \eqref{decU}, that has $\M^-$ in $\CW$, the monopole $\M^+$ \emph{extends} to $\M^{+,+}$, while the monopole $\M^-$ becomes $\M^{0,-}$. This rule will be useful to fully deconfine the theory.
We will give the full map of the chiral ring generators in \eqref{mapUT123}.

The next step is to dualize the right node $U(N)$ in $\CT_{1'}$ using Aharony duality \cite{Aharony:1997gp}
\be \ba{c} U(N) \,\, \textrm{w/} \,\, (F,F) \,\, \textrm{flavors},\\ \CW=0 \ea
 \Longleftrightarrow 
 \ba{c}U(F-N)\,\, \textrm{w/} \,\, (F,F) \,\, \textrm{flavors}\,\, p_i,\pt_i,\\ \CW= M^{ij} \tr(p_i \pt_j) + \s^\pm \M^\pm \ea \ee
in the quiver $\CT_{1'}$ and obtain $\CT_{2}$:
%
\be
\begin{tikzpicture}[baseline]
\scalebox{1.1}{
\tikzstyle{every node}=[font=\footnotesize]
\node[draw=none]at (-3.5,1.5) {$\CT_2:$};
\node[draw=none] (g1) at (-1.6,0) {$U(N-1)$};
\node[draw=none] (g2) at (1.6,0) {$U(F-1)$};
\node[draw, rectangle] (f1) at (0,2.4) {$F$};
\draw[->] (g2) to [bend left=15] (f1);
\draw[<-] (g2) to [bend right=15] (f1);
\node[draw=none] at (-1.6, 1.4) {$p, \pt$} ;
\draw[->] (g1) to [bend left=15] (f1);
\draw[<-] (g1) to [bend right=15] (f1);
\node[draw=none] at (1.6, 1.4) {$q, \qt$} ;
\draw[transform canvas={yshift=2.5pt},->] (g1) to (g2) ;
\draw[transform canvas={yshift=-2.5pt},<-] (g1) to (g2) ;
\node[draw=none] at (0, -0.4) {$b, \bt$} ;
\draw[black,solid] (g1) edge [out=-45,in=-135,loop,looseness=4] (g1);
\node[draw=none] at (-2.3, -0.6) {$\phi$} ;
\draw[black,solid] (f1) edge [out=45,in=135,loop,looseness=4] (f1);
\node[draw=none] at (0.7, 2.9) {$M$} ;
\node[draw=none] at (5,1.5) {$ \CW=  \M^{-,-} + \gamma \M^{+,+} + \s^{\pm} \M^{0,\pm} +$} ;
\node[draw=none] at (5.9,1) {$ + \tr(\bt \phi b)+ \tr(b q p) + \tr(\bt \qt \pt)  + M \tr(q \qt)$};
}
\end{tikzpicture}
\ee

We decomposed the Seiberg dual mesons into the fields $\phi$, $M$ and $p$. Because of the F-terms of the singlet $\b$, that we integrated out, the antisymmetric field $\phi$ is traceless. Notice that the monopoles $\M^{\pm,0}$ in $\CT_{1'}$ became $\M^{\pm(1,1)}$ in $\CT_{2}$, here we are applying the rules of \cite{Pasquetti:2019tix, Benvenuti:2020wpc} for the mapping of monopole operators under dualities in quivers. 

The mapping between the $R$-charges  of theories $\CT_{1}$ and $\CT_2$ is dictated by the mapping of the mesonic operators and is
\be\label{mapU1}  r_q = 1-r_F\,, \qquad r_p=r_\Phi/2+r_F\,, \qquad r_{\phi}=r_\Phi\,, \qquad r_b=1-r_\Phi/2\,.\ee

The $R$-charges of the monopoles and of the flipping fields for the monopoles are
\be\ba{rcl}
 R[\M^{+,0}]_{\CT_{2}} &=& (N-2)(1-r_\phi) + (F-1)(1-r_b) + F(1-r_p) - (N-2) + \alpha_1,  \\
 R[\M^{-,0}]_{\CT_{2}} &=& (N-2)(1-r_\phi) + (F-1)(1-r_b) + F(1-r_p) - (N-2) - \alpha_1,  \\
 R[\s^+]_{\CT_{2}} &=&2-\left( (N-1)(1-r_b) + F(1-r_q) - (F-2) +\alpha_2 \right)   \\
 R[\s^-]_{\CT_{2}} &=& 2-\left( (N-1)(1-r_b) + F(1-r_q) - (F-2) -\alpha_2 \right)  \\
 R[\gamma]_{\CT_{2}} &=& 2-( (N-2)(1-r_\phi)+(N+F-4)(1-r_b)\\
 & +& F(1-r_q) + F(1-r_p) - (N-2)- (F-2) +\alpha_1+\alpha_2) \ea\ee
where, the procedure to find $\alpha_1, \, \alpha_2$ explained in \ref{adjU}, gives
\be
	\alpha_1=-\frac{r_\Phi}{2}, \qquad \alpha_2=\frac{(1-N)\,r_\Phi}{2}.
\ee

In $\CT_{1'}$, some monopole operators can be dressed using the meson made by bifundamental fields $b$, $\bt$, as discussed in \cite{Pasquetti:2019tix, Benvenuti:2020wpc}. In $\CT_2$, some monopole operators can be dressed with the adjoint $\phi$.

The mapping of the chiral ring generators of the three theories constructed $\CT_1$, $\CT_{1'}$ and $\CT_2$ is
\be \label{mapUT123}
\ba{c}\CT_1 \\
\tr( \Qt_i Q_j)  \\
\tr( \Qt_i \Phi^J Q_j)  \\
\tr(\Phi^J)\\
\tr(\Phi^{N})\\
 \{\M^+_{\Phi^J}\}\\
 \{\M^+_{\Phi^{N-1}}\}\\
 \M^-\\
  \{\M^-_{\Phi^{J+1}}\}
\ea
\qquad \Longleftrightarrow \qquad
\ba{c} \CT_{1'}\\
\tr( \Qt_i Q_j)  \\
\tr( \Qt_i (b' \bt')^J Q_j)  \\
\tr((b' \bt')^J)\\
\tr((b' \bt')^N)\\
 \{\M^{ +,+}_{(b' \bt')^J}\}\\
 \M^{ 0,+}\\
 \M^{0,-}\\
  \{\M^{-,-}_{(b' \bt')^J}\}
\ea
\qquad \Longleftrightarrow \qquad
\ba{c}\CT_2\\
M_{ij} \\
 \tr(\pt_i \phi^{J-1} p_j)\\
\tr(\phi^J)\\
 \gamma \\
 \{\M^{+,0}_{\phi^J}\} \\
 \s^+ \\
 \s^-  \\
 \{\M^{-,0}_{\phi^J}\} 
\ea
\qquad
\ba{l}
\,\\
\,\\
J=1,\ldots,N-1\\
J=2,\ldots,N-1\\
\\
J=0,1,\ldots,N-2\\
\\
\\
J=0,1,\ldots,N-2
\,\ea
 \ee

\subsection{Fully deconfined tail}
We can proceed in a similar fashion, deconfine an adjoint field and dualize. We do not give the details since they are very similar to the $Sp$ case discussed in section \ref{asymmSp}.

After $N-1$ steps, the leftmost group is $U(1)$ so there is no adjoint traceless  to deconfine, and we just dualize the $U(1)$.

 The final result is that the starting theory $\CT_1$
\be\label{fdt}
\begin{tikzpicture}[baseline]
\scalebox{1.1}{
\tikzstyle{every node}=[font=\footnotesize]
\node[draw=none]at (-2,0) {$\CT_1:$};
\node[draw=none] (g1) at (0,0) {$U(N)$};
\node[draw, rectangle] (f1) at (2.5,0) {$F$};
\draw[transform canvas={yshift=2.5pt},->] (g1) to (f1) ;
\draw[transform canvas={yshift=-2.5pt},<-] (g1) to (f1) ;
\node[draw=none] at (1.3, -0.5) {$Q, \Qt$} ;
\draw[black,solid] (g1) edge [out=45,in=135,loop,looseness=4] (g1);
\node[draw=none] at (0, 1.2) {$\Phi$} ;
\node[draw=none] at (4.5, 0) {$\CW=0$} ;
}
\end{tikzpicture}
\ee
is dual to a quiver with $N$ gauge nodes $\CT_{DEC}$
\be\label{TDECU}
\begin{tikzpicture}[baseline]
\scalebox{1}{
\tikzstyle{every node}=[font=\footnotesize]
\node[draw=none]at (-3-1,1.5) {$\CT_{DEC}:$};
\node[draw=none] (g1) at (0-1,0) {$U(N(F-1))$};
\node[draw=none] (g2) at (4-1,0) {$U((N-1)(F-1))$};
\node[draw=none] (g3) at (7-1,0) {$\cdots$};
\node[draw=none] (g4) at (9.3-1,0) {$U(F-1)$};
\node[draw, rectangle] (f1) at (0-1,2.5) {$F$};
\draw[black,solid] (g2) edge [out=-45,in=-135,loop,looseness=4] (g2);
\node[draw=none] at (4.1-1, -1.2) {$\Phi_{N-1}$} ;
\draw[black,solid] (g4) edge [out=-45,in=-135,loop,looseness=4] (g4);
\node[draw=none] at (9.4-1, -1.2) {$\Phi_{1}$} ;
\draw[transform canvas={yshift=2.5pt},->] (g1) to (g2) ;
\draw[transform canvas={yshift=-2.5pt},<-] (g1) to (g2) ;
\node[draw=none] at (1.8-1, -0.5) {$b_{N-1}, \bt_{N-1}$} ;
\draw[transform canvas={yshift=2.5pt},->] (g2) to (g3) ;
\draw[transform canvas={yshift=-2.5pt},<-] (g2) to (g3) ;
\node[draw=none] at (6.2-1, -0.5) {$b_{N-2}, \bt_{N-2}$} ;
\draw[transform canvas={yshift=2.5pt},->] (g3) to (g4) ;
\draw[transform canvas={yshift=-2.5pt},<-] (g3) to (g4) ;
\node[draw=none] at (8-1, -0.5) {$b_{1}, \bt_{1}$} ;
\draw[transform canvas={xshift=2.5pt},->] (g1) to (f1) ;
\draw[transform canvas={xshift=-2.5pt},<-] (g1) to (f1) ;
\node[draw=none] at (0.5-1, 1.2) {$p, \pt$} ;
\draw[black,solid] (f1) edge [out=45,in=135,loop,looseness=2] (f1);
\draw[black,solid] (f1) edge [out=45,in=135,loop,looseness=3] (f1);
\draw[black,solid] (f1) edge [out=45,in=135,loop,looseness=7] (f1);
\node[draw=none] at (0.03-1, 3.25) {$\cdots$} ;
\node[draw=none] at (0.9-1, 3.25) {$M_i$} ;
\node[draw=none] at (4.3-1, 3.3)  {$ \CW= \sum_{i=1}^{N-1}  \M^{0^{N-i}, - , 0^{i-1}} + $};
\node[draw=none] at (6.6-1, 2.7)  {$ + \sum_{i=1}^{N-1} \g_i \M^{0, +^{N-i}, 0^{i-1}} +\sum_{i=1}^{N} \s^{\pm}_i \M^{\pm^{N-i+1}, 0^{i-1}}  + $};
\node[draw=none] at (5.72-1, 2.1)  {$  +  \sum_{i=1}^{N} M_i \tr(\pt \bt_{N-1} \ldots \bt_{i} b_{i}  \ldots b_{N-1} p) +$};
\node[draw=none] at (6.49-1, 1.5) {$ + \sum_{i=1}^{N-1} ( \tr(\Phi_i \bt_i b_i) + \tr(\Phi_i \bt_{i-1} b_{i-1}) + \phi_i \tr(\bt_i b_i)) $};
}
\end{tikzpicture}
\ee

The $R$-charges of the elementary fields in the fully deconfined theory are given in terms of the two independent $R$-charges in theory $1$, $r_F$ and $r_\Phi$ as follows:
\be \ba{l}
r_{b_i} = r_\Phi/2 \qquad i=1,\ldots,N-1 \\
r_{\Phi_i} = r_{\phi_i}  = 2-r_\Phi  \qquad i=1,\ldots,N-1\\
r_p = 1- (N-1) r_\Phi/2 - r_F\\
r_{M_i} =2 r_F + (i-1) r_\Phi  \qquad i=1,\ldots,N
\ea \ee
The mapping of the chiral ring generators in the duality $\CT_1 \leftrightarrow \CT_{DEC}$ is
\be \label{mapt1tdecU}
\ba{c}\CT_1 \\
\tr( \Qt_i \Phi^J Q_j)  \\
\tr(\Phi^J)\\
 \{\M^+_{\Phi^{J}}\}\\
 \{\M^-_{\Phi^{J}}\}
  \ea
\qquad \Longleftrightarrow \qquad
\ba{c} \CT_{DEC} \\
 (M_{J+1})_{ij}  \\
\g_{N-J+1}\\
\s^+_{N-J}\\
\s^-_{J+1}
 \ea
\quad\qquad
\ba{l}
\,\\
J=0,\ldots,N-1\\
J=2,\ldots,N\\
J=0,\ldots,N-1\\
J=0,\ldots,N-1
  \ea
 \ee
 
Notice that $ \{\M^+_{\Phi^{J}}\}$ monopoles map to singlets $\s^+_{N-J}$, in the same way of the monopoles of $Sp(N)$ with an antisymmetric, \eqref{mapt1tdec}. On the other hand $ \{\M^-_{\Phi^{J}}\}$ monopoles map to $\s^-_{J+1}$. This is due to the fact that every time we deconfine the rank-2 field, the positive charge monopoles of $U$ and the monopoles of $Sp$ extend ($\M^{+,\ldots}$ becomes $\M^{+,+,\ldots}$, $\M^{\cdot,\ldots}$ becomes $\M^{\cdot,\cdot,\ldots}$), while the negative charge monopoles of $U$ do not extend ($\M^{-,\ldots}$ becomes $\M^{0,-,\ldots}$).

The general formula for the monopole $R$-charge in $\CT_{DEC}$ reads
\be
	R[\mathfrak M^{m^{(1)}, m^{(2)}, \cdots, m^{(N)}}]=(\text{standard}) \,+ \,\alpha_1 \sum_{i_1=1}^{N(F-1)} m^{(1)}_{i_1}+\cdots+ \,\alpha_N \sum_{i_N=1}^{F-1} m^{(N)}_{i_N},
\ee
where
\be
	\alpha_1=\frac{N-1}{2}\,r_{\Phi}, \qquad \alpha_2=\cdots=\alpha_N=-r_{\Phi}.
\ee
Observe that the superpotential for $\CT_{DEC}$ contains $N-1$ linear monopoles, and their marginality fixes $N-1$ of the $\alpha_i$ parameters; the remaining one has to be fixed using the duality map. 

Let us finally comment that, as for Aharony duality for $U$ gauge group without rank-$2$ matter fields, all the chiral ring generators of $\CT_1$ map to gauge singlets in $\CT_{DEC}$.

\subsection{Superpotential deformation: $\CW= \text{tr} (\Qt_{F} \Phi^J Q_{F})$}\label{seccdef1}
In this section and in section \ref{sec:N=4U}, we discuss complex deformations of the duality between our original theory with a single node $U(N)$  $\CT_1$ \eqref{T1U} and the fully deconfined quiver $\CT_{DEC}$ \eqref{TDECU}. As usual in Seiberg-like dualities, a complex deformation on the electric side will induce a Higgsing of the gauge groups on the magnetic side.

In this section we consider a superpotential deformation in $\CT_1$ of the form $\tr(\tilde{Q}_{F} \Phi^J Q_{F})$. This meson, according to \eqref{mapt1tdecU}, is mapped in $\CT_{DEC}$ to  $(M_{J+1})_{F,F}$ , the flipping field for the meson $\tr(\tilde{p}_{F} \ldots \tilde{b}_{J+1} b_{J+1}  \ldots p_{F})$. We take $J<N$. Turning on a linear superpotential term in $M_J$ means that this $F \times F$ matrix of long mesons must take a non-zero vacuum expectation value of minimal non zero rank. This is achieved by giving a vev to the bifundamentals $b_{J+1}, \ldots, b_{N-1}$ and to the flavors $p$, such that $\tr(\tilde{p}_{F}\tilde{b}_{N-1} \ldots \tilde{b}_{J+1} b_{J+1}  \ldots b_{N-1} p_F)=1$, while all other mesons are zero. 

 Without going too much into the details, the final result is that in \eqref{TDECU} the $N-J$ gauge groups on the left
 $$U(N(F-1)),\ldots, U(h(F-1)), \ldots, U((J+1)(F-1))$$ 
 are Higgsed to 
 $$U(N(F-1)-(N-J)),\ldots, U(h(F-1)-(h-J)), \ldots, U((J+1)(F-1)-(J+1-J))\,,$$
while the $J$ remaining gauge groups on the right are not Higgsed. The last flavor $\tilde{p}_{F},p_{F}$ \emph{migrates} from the left-most node to node $U(J(F-1))$. More precisely, since the node $U((J+1)(F-1))$ is Higgsed down to $U((J+1)(F-2)+J)$, the bifundamental field $b_{J}$ splits into a new $U((J+1)(F-2)+J) - U(J(F-1))$ bifundamental and a flavor for the node $U(J(F-1))$. The flipping fields for the mesons split into two sets $M_H$, $H=1,\ldots,N$, and $(M')_K$, $K=1,\ldots,J$.

 The final result is that 
\be
\begin{tikzpicture}[baseline]
\scalebox{1.1}{
\tikzstyle{every node}=[font=\footnotesize]
\node[draw=none] (g1) at (0,0) {$U(N)$};
\node[draw, rectangle] (f1) at (2.5,0) {$F$};
\draw[transform canvas={yshift=2.5pt},->] (g1) to (f1) ;
\draw[transform canvas={yshift=-2.5pt},<-] (g1) to (f1) ;
\node[draw=none] at (1.3, -0.5) {$Q, \Qt$} ;
\draw[black,solid] (g1) edge [out=45,in=135,loop,looseness=4] (g1);
\node[draw=none] at (0, 1.2) {$\Phi$} ;
\node[draw=none] at (6, 0) {$\CW= \text{tr}(\Qt_{F} \Phi^J Q_F)$} ;
}
\end{tikzpicture}
\ee
is dual to
\be
\begin{tikzpicture}[baseline]
\scalebox{1}{
\tikzstyle{every node}=[font=\footnotesize]
\node[draw=none] (g1) at (0,0) {$U(N(F-2)+J)$};
\node[draw=none] (g2) at (4,0) {$U((N-1)(F-2)+J)$};
\node[draw=none] (g3) at (7,0) {$\cdots$};
\node[draw=none] (g4) at (9.3,0) {$U(J(F-1))$};
\node[draw=none] (g5) at (11.5,0) {$\cdots$};
\node[draw=none] (g6) at (13.5,0) {$U(F-1)$};
\node[draw, rectangle] (f1) at (0,2.5) {$F-1$};
\node[draw, rectangle] (f2) at (9.3,2.5) {$1$};
\draw[black,solid] (g2) edge [out=-45,in=-135,loop,looseness=4] (g2);
\node[draw=none] at (4.1, -1.2) {$\Phi_{N-1}$} ;
\draw[black,solid] (g4) edge [out=-45,in=-135,loop,looseness=4] (g4);
\node[draw=none] at (9.4, -1.2) {$\Phi_{J}$} ;
\draw[black,solid] (g6) edge [out=-45,in=-135,loop,looseness=4] (g6);
\node[draw=none] at (13.6, -1.2) {$\Phi_{1}$} ;
\draw[transform canvas={yshift=2.5pt},->] (g1) to (g2) ;
\draw[transform canvas={yshift=-2.5pt},<-] (g1) to (g2) ;
\node[draw=none] at (1.8, -0.5) {$b_{N-1}$} ;
\draw[transform canvas={yshift=2.5pt},->] (g2) to (g3) ;
\draw[transform canvas={yshift=-2.5pt},<-] (g2) to (g3) ;
\node[draw=none] at (6.2, -0.5) {$b_{N-2}$} ;
\draw[transform canvas={yshift=2.5pt},->] (g3) to (g4) ;
\draw[transform canvas={yshift=-2.5pt},<-] (g3) to (g4) ;
\node[draw=none] at (7.9, -0.5) {$b_{J}$} ;
\draw[transform canvas={yshift=2.5pt},->] (g4) to (g5) ;
\draw[transform canvas={yshift=-2.5pt},<-] (g4) to (g5) ;
\node[draw=none] at (10.8, -0.5) {$b_{J-1}$} ;
\draw[transform canvas={yshift=2.5pt},->] (g5) to (g6) ;
\draw[transform canvas={yshift=-2.5pt},<-] (g5) to (g6) ;
\node[draw=none] at (12.3, -0.5) {$b_{1}$} ;
\draw[transform canvas={xshift=2.5pt},->] (g1) to (f1) ;
\draw[transform canvas={xshift=-2.5pt},<-] (g1) to (f1) ;
\draw[transform canvas={xshift=2.5pt},->] (g4) to (f2) ;
\draw[transform canvas={xshift=-2.5pt},<-] (g4) to (f2) ;
\draw[black,solid] (f1) edge [out=45,in=135,loop,looseness=2] (f1);
\draw[black,solid] (f1) edge [out=45,in=135,loop,looseness=3] (f1);
\draw[black,solid] (f1) edge [out=45,in=135,loop,looseness=7] (f1);
\node[draw=none] at (0.03, 3.25) {$\cdots$} ;
\node[draw=none] at (-0.8, 3.25) {$M_i$} ;
\node[draw=none] at (0.35, 1.5) {$p$} ;
\node[draw=none] at (9.7, 1.5) {$p'$} ;
\draw[black,solid] (f2) edge [out=45,in=135,loop,looseness=2] (f2);
\draw[black,solid] (f2) edge [out=45,in=135,loop,looseness=3] (f2);
\draw[black,solid] (f2) edge [out=45,in=135,loop,looseness=7] (f2);
\node[draw=none] at (9.32, 3.15) {$\cdots$} ;
\node[draw=none] at (10.1, 3.2) {$M'_i$} ;
\node[draw=none] at (6.4, 4.2) {$ \CW=  \sum_{i=1}^{N-1}  \M^{0^{N-i}, - , 0^{i-1}} +\sum_{i=1}^{N-1} \g_i \M^{0, +^{N-i}, 0^{i-1}} +\sum_{i=1}^{N} \s^{\pm}_i \M^{\pm^{N-i+1}, 0^{i-1}}  +$};
\node[draw=none] at (4.9, 3.5)  {$ + \sum_{i=1}^{N-1} ( \tr(\Phi_i \bt_i b_i) + \tr(\Phi_i \bt_{i-1} b_{i-1}) + \phi_i \tr(\bt_i b_i))+$};
\node[draw=none] at (4, 2.8){$  +  \sum_{i=1}^{N} M_i \tr(\tilde{p} \tilde{b}_{N-1} \ldots \tilde{b}_{i} b_{i}  \ldots b_{N-1} p) + $};
\node[draw=none] at (3.9, 2.1){$ + \sum_{i=1}^{J} M'_i \tr(\tilde{p}' \tilde{b}_{J-1} \ldots \tilde{b}_{i} b_{i}  \ldots b_{J-1} p') $} ;
}
\end{tikzpicture}
\ee
 Turning on a complex mass for a single flavor is a special case  $J=0$ of the discussion above. The flavors $p'$ and the gauge singlets $M'$ are absent for $J=0$, and all the gauge groups in $\CT_{DEC}$ get partially Higgsed. The final result is precisely \eqref{TDECU}  with $F \rightarrow F-1$. 

\subsection{Deformation to the \emph{$\CN\!=\!4$} theory: $ \CW=\sum_{j=1}^F \tr(\Qt_{j} \Phi Q_j)$}\label{sec:N=4U}
We now  add $f$ cubic terms to $\CT_1$, obtaining $U(N)$ with $F$  flavor hypers and $ \CW=\sum_{j=1,\ldots,F} \tr(\Qt_{j} \Phi Q_j)$, that is the $\CN=4$ theory $U(N)$ with $F$ flavors and flavor symmetry $SU(F) \times U(1)_{\text{top}}$.\footnote{The precise value of the coupling constant $\lambda$ in the superpotential  $\cW= \lambda \sum_{j=1,\ldots,F} \tr(\Qt_{j} \Phi Q_j)$ is not crucial for the claim that the RG flow triggered by $\lambda$ lands on the $\cN=4$ theory. In the $2d$ space of the gauge coupling and $\lambda$, there is only one fixed point with non-zero gauge coupling and non-zero $\lambda$, namely the $\cN=4$ SCFT.} 

Using the results obtained in Section \ref{seccdef1}, on the magnetic side all the flavors \emph{migrate} to the right-most node $U(F-1)$, and out of the tower of singlets $(M_J)_{ij}$, only $(M_1)_{ij}$ survive. The tail of gauge groups 
$$U(N(F-1)),\ldots,U(3(F-1)), U(2(F-1)), U(F-1)$$ 
Higgses to 
$$U(F-N),\ldots,U(F-3), U(F-2), U(F-1)\,.$$ 
The right-most  group $U(F-1)$ is not Higgsed.

The final result is that 
\be
\begin{tikzpicture}[baseline]
\scalebox{1.1}{
\tikzstyle{every node}=[font=\footnotesize]
\node[draw=none] at (-2,0) {$\CT_1:$};
\node[draw=none] (g1) at (0,0) {$U(N)$};
\node[draw, rectangle] (f1) at (2.5,0) {$F$};
\draw[transform canvas={yshift=2.5pt},->] (g1) to (f1) ;
\draw[transform canvas={yshift=-2.5pt},<-] (g1) to (f1) ;
\node[draw=none] at (1.3, -0.5) {$Q, \Qt$} ;
\draw[black,solid] (g1) edge [out=45,in=135,loop,looseness=4] (g1);
\node[draw=none] at (0, 1.2) {$\Phi$} ;
\node[draw=none] at (6, 0) {$\CW=\sum_{j=1}^{F} \text{tr}(\Qt_{j} \Phi^J Q_j)$} ;
}
\end{tikzpicture}
\ee
is dual to
\be
\begin{tikzpicture}[baseline]
\scalebox{1}{
\tikzstyle{every node}=[font=\footnotesize]
\node[draw=none] (g1) at (0-1,0) {$U(F-N)$};
\node[draw=none] (g2) at (3.5-1,0) {$U(F-N+1)$};
\node[draw=none] (g3) at (6.5-1,0) {$\cdots$};
\node[draw=none] (g4) at (9.3-1,0) {$U(F-1)$};
\node[draw, rectangle] (f1) at (9.3-1,2.5) {$F$};
\draw[black,solid] (g2) edge [out=-45,in=-135,loop,looseness=4] (g2);
\node[draw=none] at (3.5-1, -1.2) {$\Phi_{N-1}$} ;
\draw[black,solid] (g4) edge [out=-45,in=-135,loop,looseness=4] (g4);
\node[draw=none] at (9.4-1, -1.2) {$\Phi_{1}$} ;
\draw[transform canvas={yshift=2.5pt},->] (g1) to (g2) ;
\draw[transform canvas={yshift=-2.5pt},<-] (g1) to (g2) ;
\node[draw=none] at (1.8-1, -0.5) {$b_{N-1}, \bt_{N-1}$} ;
\draw[transform canvas={yshift=2.5pt},->] (g2) to (g3) ;
\draw[transform canvas={yshift=-2.5pt},<-] (g2) to (g3) ;
\node[draw=none] at (5.5-1, -0.5) {$b_{N-2}, \bt_{N-2}$} ;
\draw[transform canvas={yshift=2.5pt},->] (g3) to (g4) ;
\draw[transform canvas={yshift=-2.5pt},<-] (g3) to (g4) ;
\node[draw=none] at (8-1, -0.5) {$b_{1}, \bt_{1}$} ;
\draw[transform canvas={xshift=2.5pt},->] (g4) to (f1) ;
\draw[transform canvas={xshift=-2.5pt},<-] (g4) to (f1) ;
\node[draw=none] at (9.8-1, 1.3) {$p, \pt$} ;
\draw[black,solid] (f1) edge [out=45,in=135,loop,looseness=6] (f1);
\node[draw=none] at (9.1, 3.15) {$M_i$} ;
\node[draw=none] at (3, 3){$ \CW=  \sum_{i=1}^{N-1}  \M^{0^{N-i}, - , 0^{i-1}} +\sum_{i=1}^{N-1} \g_i \M^{0, +^{N-i}, 0^{i-1}}  + $};
\node[draw=none] at (2.4, 2.3) {$ +\sum_{i=1}^{N} \s^{\pm}_i \M^{\pm^{N-i+1}, 0^{i-1}}  + M \tr(p \tilde p) +$};
\node[draw=none] at (3.3, 1.6) {$ + \sum_{i=1}^{N-1} ( \text{tr}(\Phi_i \bt_i b_i) + \text{tr}(\Phi_i \bt_{i-1} b_{i-1}) + \phi_i \text{tr}(\bt_i b_i)) $};
}
\end{tikzpicture}
\ee

%
This result is strictly speaking valid for $F>N$. If $F\leq N$ the dual quiver becomes shorter and some of the flipping fields $\g$ and $\s$ decouple. This is due to the fact the theory $U(N)$ with $\CW= \sum_{j=1,\ldots,F} \text{tr}(\tilde{Q}_{j} \Phi Q_j)$ if $F \leq N$ becomes ``bad'' in the Gaiotto-Witten sense, so some Coluomb branch operators (that is $\tr(\Phi^h)$ and $\{\M_{\Phi^k}\}$) become free and decouple.

\subsection{Real masses and Chern-Simons terms}\label{sec:addCSU}
It is immediate to start from the duality between \eqref{fdt} and the fully deconfined tail \eqref{TDECU} and derive the corresponding duality in the presence of a non-trivial Chern-Simons level. There are various possibilities to generate a CS level, and our aim in this section is only to give one example and not to treat the most general case, as for instance it has been done in \cite{Benini:2011mf} in the case without adjoint matter. The example we focus is as follows. We start from \eqref{fdt} with $F+k$ flavours and give a real mass to $k$ of the fundamental chirals $Q$. The result on the electric side is
\be
\begin{tikzpicture}[baseline]
\scalebox{1.1}{
\tikzstyle{every node}=[font=\footnotesize]
\node[draw=none] at (-2,0) {$\CT_1^{\text{CS}}:$};
\node[draw=none] (g1) at (0,0) {$U(N)_{\tfrac{k}{2}}$};
\node[draw, rectangle] (f1) at (2.5,0.8) {$F$};
\node[draw, rectangle] (f2) at (2.5,-0.8) {$F+k$};
\draw[black,solid,->] (g1)--(f1);
\draw[black,solid,<-] (g1)--(f2);
\node[draw=none] at (1.5, 0.8) {$Q$} ;
\node[draw=none] at (1.5, -0.8) {$\Qt$} ;
\draw[black,solid] (g1) edge [out=45,in=135,loop,looseness=4] (g1);
\node[draw=none] at (-0.9, 0.8) {$\Phi$} ;
\node[draw=none] at (4.5, 0) {$\CW=0$} ;
}
\end{tikzpicture}
\ee
The effect of having a CS term is to remove some of the monopoles from the chiral ring. In general, the fundamental monopole operators $\mathfrak M^{\pm}$ acquire a gauge charge under the $U(1)$ part of the gauge group given by
\be
	\mp \left[k_{\text{CS}} \pm \frac{1}{2}(N_f-N_a) \right] = 
	\mp \left[ \frac{k}{2} \pm \frac{1}{2}(F-(F+k)) \right] =
	\begin{cases}
		0 \quad \quad\text{for} \; \mathfrak M^{+} \\ 
		-k \quad \,\text{for} \; \mathfrak M^{-} \\ 
	\end{cases}
\ee
thus, the monopoles negatively charged under the topological symmetry are removed from the chiral ring since gauge variant. 

The dual is:\footnote{As in footnote \ref{FNlowr}, it is interesting to consider special case with low $F$ and $k$. The story is similar.

If $F+|k|=1$, all the ranks in the quiver tails vanish. In the case of the fully deconfined theory, the full gauge group is trivial. This means that the deconfined theory is replaced by a Wess-Zumino, with a non trivial superpotential constructed out of the gauge singlet fields $\g_i, \s_i, M_i$, as in  \cite{Amariti:2018wht, Benvenuti:2018bav}. 

If $F+|k|=2$, the quiver tail is $U(N)_{-k} - U(N-1)- U(N-2) -\ldots - U(1)$, and this tail can be \emph{sequentially confined}. We start from the right-most node, which is a $U(1)_0$ with $(2,2)$ flavors, so it confines. Moreover, the adjoint for the $U(2)_0$ node is removed. We then dualize the $U(2)_0$ with $(3,3)$ flavors, which also confines. After $N-1$ confining steps, we end up with $U(N)_{-k}$, with an adjoint plus $(2-|k|,2)$ flavors, and possibly some flipping fields. The same $U(N)_{+k} \leftrightarrow U(N)_{-k}$ duality can be achieved in a different way, for instance turning on some real masses in duality 2.34 of \cite{Benvenuti:2018bav}.}

\be
\begin{tikzpicture}[baseline]
\scalebox{1}{
\tikzstyle{every node}=[font=\footnotesize]
\node[draw=none]at (-2.5-1,1.5) {$\CT_{\text{DEC}}^{\text{CS}}:$};
\node[draw=none] (g1) at (0-1,0) {$U(N(F+k-1))_{-\tfrac{k}{2}}$};
\node[draw=none] (g2) at (4.5-1,0) {$U((N-1)(F+k-1))$};
\node[draw=none] (g3) at (8-1,0) {$\cdots$};
\node[draw=none] (g4) at (10.7-1,0) {$U(F+k-1)$};
\node[draw, rectangle] (f1) at (-1-1,2.5) {$F$};
\node[draw, rectangle] (f2) at (1-1,2.5) {$F+k$};
\draw[black,solid] (g2) edge [out=-45,in=-135,loop,looseness=4] (g2);
\node[draw=none] at (4.6-1, -1.2) {$\Phi_{N-1}$} ;
\draw[black,solid] (g4) edge [out=-45,in=-135,loop,looseness=4] (g4);
\node[draw=none] at (10.8-1, -1.2) {$\Phi_{1}$} ;
\draw[transform canvas={yshift=2.5pt},->] (g1) to (g2) ;
\draw[transform canvas={yshift=-2.5pt},<-] (g1) to (g2) ;
\node[draw=none] at (2.3-1, -0.5) {$b_{N-1}, \bt_{N-1}$} ;
\draw[transform canvas={yshift=2.5pt},->] (g2) to (g3) ;
\draw[transform canvas={yshift=-2.5pt},<-] (g2) to (g3) ;
\node[draw=none] at (7-1, -0.5) {$b_{N-2}, \bt_{N-2}$} ;
\draw[transform canvas={yshift=2.5pt},->] (g3) to (g4) ;
\draw[transform canvas={yshift=-2.5pt},<-] (g3) to (g4) ;
\node[draw=none] at (9-1, -0.5) {$b_{1}, \bt_{1}$} ;
\draw[black,solid,<-] (g1) to (f1) ;
\node[draw=none] at (-0.8-1, 1.2) {$\pt$} ;
\draw[black,solid,->] (g1) to (f2) ;
\node[draw=none] at (0.8-1, 1.2) {$p$} ;
\node[draw=none] at (4.3-1, 3.3)  {$ \CW= \sum_{i=1}^{N-1}  \M^{0^{N-i}, - , 0^{i-1}} + $};
\node[draw=none] at (6.6-1, 2.7)  {$ + \sum_{i=1}^{N-1} \g_i \M^{0, +^{N-i}, 0^{i-1}} +\sum_{i=1}^{N} \s^{+}_i \M^{+^{N-i+1}, 0^{i-1}}  + $};
\node[draw=none] at (5.72-1, 2.1)  {$  +  \sum_{i=1}^{N} M_i \tr(\pt \bt_{N-1} \ldots \bt_{i} b_{i}  \ldots b_{N-1} p) +$};
\node[draw=none] at (6.49-1, 1.5) {$ + \sum_{i=1}^{N-1} ( \tr(\Phi_i \bt_i b_i) + \tr(\Phi_i \bt_{i-1} b_{i-1}) + \phi_i \tr(\bt_i b_i)) $};
}
\end{tikzpicture}
\ee
where observe that, similarly to the electric theory, all the monopoles under flux $-$ under the node with CS level are not gauge invariant and disappear from the superpotential, correspondingly, all the $\sigma_i^-$ are removed from the chiral ring (recall that for vanishing CS level these singlets map to the tower of negatively charged dressed monopoles in the electric theory).

\acknowledgments
We are grateful to Antonio Amariti, Marco Fazzi, Simone Giacomelli, Noppadol Mekareeya, Sara Pasquetti and Matteo Sacchi for useful discussions. GLM is supported by the Swedish Research Council grant number 2015-05333 and partially supported by ERC Consolidator Grant number 772408 ``String landscape''.

\appendix

\bibliographystyle{ytphys}
\bibliography{ref}
\end{document}